\documentclass[a4paper,fleqn,usenatbib,useAMS]{mnras}
\usepackage{graphics}
\usepackage{graphicx}
\usepackage{amsmath}    
\usepackage{amssymb}    
\usepackage{multicol}        
\usepackage{bm}     
\usepackage{pdflscape}  
\usepackage{xcolor}
\usepackage{ae,aecompl}
\usepackage{subfigure}
\usepackage[english]{babel}
\usepackage{times}

\title[Polarization in binary microlensing]{Polarization in caustic-crossing binary microlensing events}
\author[Sajadian]{Sedighe Sajadian$~^{1,2}$\thanks{E-mail: s.sajadian@cc.iut.ac.ir}\\
$^{1}$Department~of~Physics,~Isfahan~University~of~Technology,~Isfahan~84156-83111,~Iran\\
$^{2}$ICRANet-Isfahan,~Isfahan~University~of~Technology,~Isfahan,~84156-83111,~Iran}

\begin{document}
\label{firstpage}
\pagerange{\pageref{firstpage}--\pageref{lastpage}}
\maketitle
\begin{abstract}
Here, we revisit the polarization signal in caustic-crossing binary
microlensing and introduce the maps of the polarization signals of
the source star as a function of its position projected on the lens
plane. The behavior of these maps depends on the source size in
comparison with the caustic. If the source size is so smaller than
the caustic curve, the maximum polarization signal occurs while the
source edge is crossing the nearest folds to the position of the
primary. When the source size is larger than the caustic, the
polarization signal maximizes over a circular ring around the
primary whose radius normalized to the source radius is
$\simeq0.96$. In cusp caustic crossings, the polarization curves
have three peaks, the largest and widest one happens when the source
edge is on the corner of the cusp and the source center is inside
the caustic curve. When the source is entirely inside the caustic,
the polarization signal significantly decreases. Despite the low
magnification factor between two parts of planetary caustics, its
polarization signal is considerable and can reach to $0.2\%$ for
early-type stars. While crossing the connection line between
different parts of caustic, the polarimetry curve have three close
peaks that the middle one appears when the source center is upon the
connection line.
\end{abstract}


\section{Introduction}
Polarization of a light occurs when its electric field oscillates in
a special plane instead of vibrating in all planes normal to the
direction of its propagation. If an unpolarized light passes through
a fluid, it will be partially or completely polarized owing to
scattering processes \citep{chandrasekhar60}. If the size of the
particles in the fluid is smaller than or in the order of the
incoming wavelength, the state of material does not change, the
so-called Rayleigh scattering, \citep[see, e.g.,][]{young1981}. In
this scattering process, the bound electron of a particle absorbs
the energy of the interacting electromagnetic waves, but this energy
is not sufficient to excite the electron. Hence, it vibrates
parallel with the electric field direction (at the time of the
collision) and propagates a light normal to its vibration direction.
The scattered light will be polarized and its polarization value
depends on the observer's line of sight. Another type of scattering
is Thomson scattering which is the elastic scattering of the
electromagnetic radiation by a free and charged particle. The
resulted radiation will be linearly polarized \citep[see,
e.g.,][]{ThomsonS}.

Both kinds of scattering take place in the stellar atmospheres.
Their contributions depend on the stellar surface temperatures and
the scattering species. In the stellar atmospheres, every point of
the stellar surface has a different polarization value. At the
center, there is no net polarization and at its edges the
polarization maximizes. Generally, it depends on the angle between
the observer's line of sight and the normal to the stellar surface.
If the source star is very far from us, we receive the integrated
light of the source star which has no net polarization, because of
the circular symmetry of the source surface. If this symmetry
breaks, a net polarization signal is measurable. Gravitational
microlensing breaks the circular symmetry of the source surface and
causes a net polarization signal
\citep{schneider1987,simmons1995a,simmons1995b,bogdanov1996}.
\begin{figure}
\begin{center}
\includegraphics[angle=0,width=0.49\textwidth,clip=0]{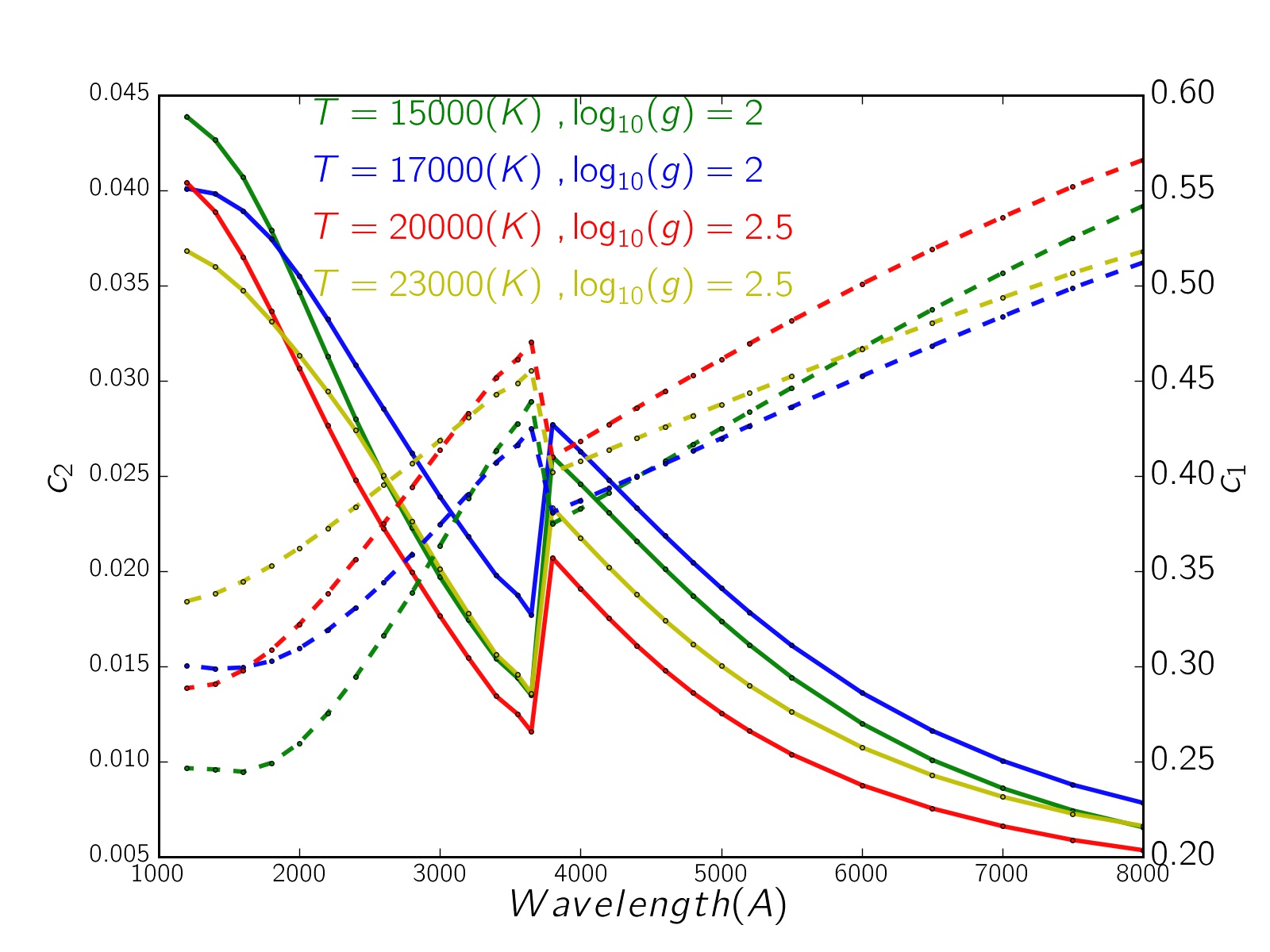}
\caption{The limb-darkening coefficients $c_{2}$ (solid curves) and
$c_{1}$ (dashed curves) for four models of early-type stars versus
the monochromatic wavelength, based on the stellar polarization
models which were developed by \citet{Harrington15}.}\label{hotstar}
\end{center}
\end{figure}

The polarization in microlensing events were studied in many
references as a helpful tool to partially resolve the degeneracy in
the microlensing parameters by measuring the source radius, the
Einstein radius, etc. \citep[see,
e.g.,][]{agol1996,yoshida2006,simmons2002}. In microlensing events,
the intrinsic polarization signals of the source stars which are
mostly unmeasurable can be magnified and generate some perturbations
in polarization curves. The intrinsic polarization signals are made
by second-order perturbations, for instance the stellar magnetic
fields and spots, circumstellar disks, close-in giant planets, etc.,
\citep[ e.g.,][]{sajadian2015,sajadian2015b,sajadian2017}.

Polarization in binary microlensing events and in the fold caustic
crossings was first analytically estimated by \citet{schneider1987}.
Then, \citet{agol1996} studied this subject and he noticed that the
polarization signal of a lensed early-type source star during
caustic crossings can reach to one percent. He discussed on the
benefits of measuring the polarization signals during binary
microlensing events of early-type source stars. The polarization
curves for different types of source stars, several detected
planetary microlensing events and the reported catalogue of single
microlensing events by the Optical Gravitational Lensing Experiment
collaboration (OGLE-III)\citep{OGLEIII} were well investigated by
\citet{Ingrosso2012,Ingrosso2015}.

Here, we revisit the polarization in caustic-crossing microlensing
events and introduce the maps of the polarization signals over the
binary lens plane and study them. We consider one type of stars,
i.e., hot and early-type stars, as the source stars and probe the
source locations where the polarization signals maximize while
passing from different caustic configurations. This study will be
beneficial for potentially follow-up polarimetry observations of
on-going binary and caustic-crossing microlensing events in near
future. It helps to recognize the best candidates of on-going
microlensing events for polarimetry observations. We expect to have
similar polarization behaviors while caustic crossings for other
types of source stars, e.g., main-sequence and red clump giant (RCG)
stars, but with different scales. Indeed, the local and intrinsic
polarization profiles over different types of stars are similar,
i.e., they depends just on the source radius projected on sky plane
and enhances while moving from the source center to its edges. We
study the polarization maps and polarization curves in the fold and
cusp caustic crossings, detailed in the section (\ref{two}). We
summarize the conclusions in the last section.

\begin{figure*}
\centering
\subfigure[]{\includegraphics[angle=0,width=0.495\textwidth,clip=0]{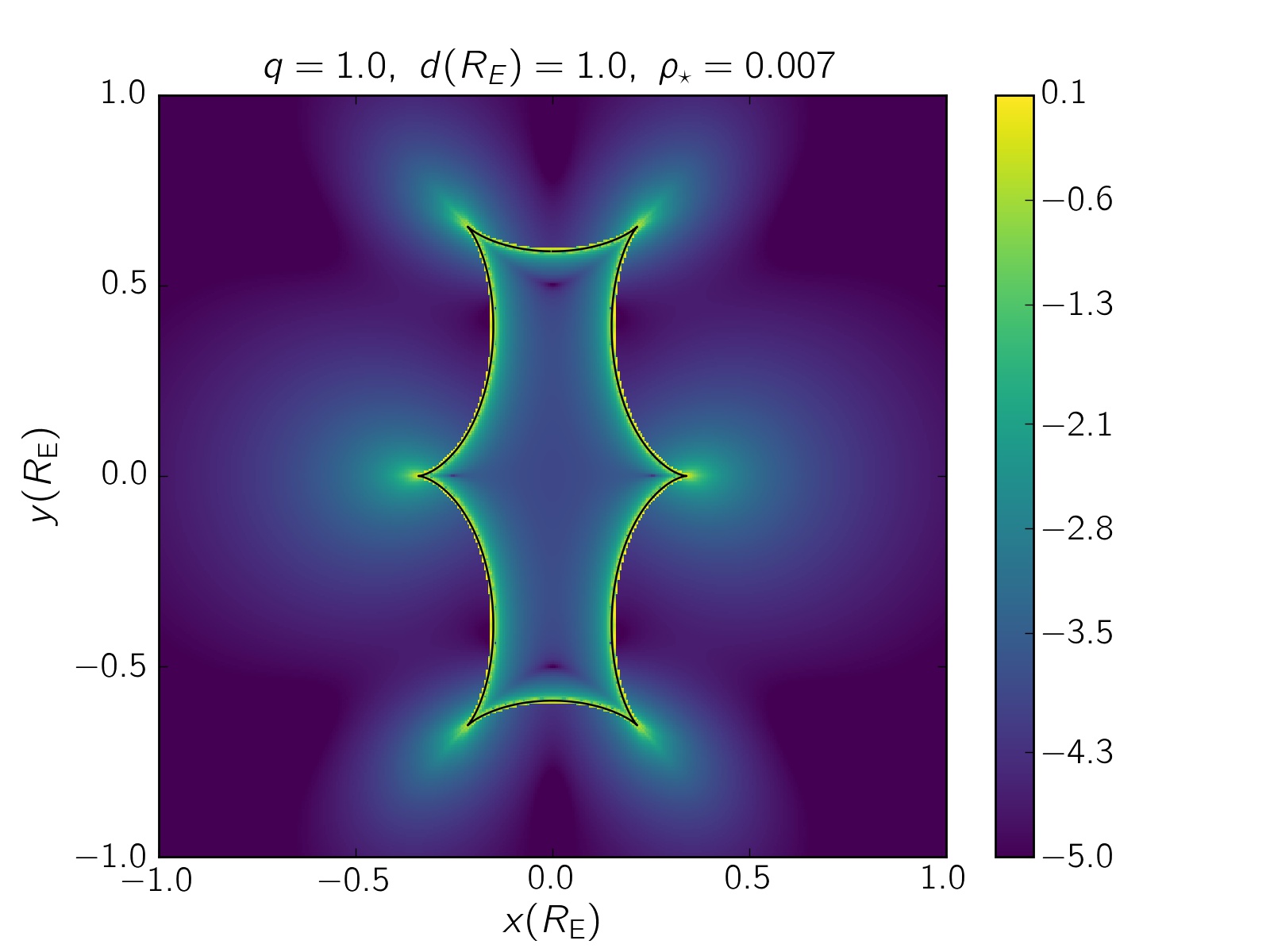}\label{fig1a}}
\subfigure[]{\includegraphics[angle=0,width=0.495\textwidth,clip=0]{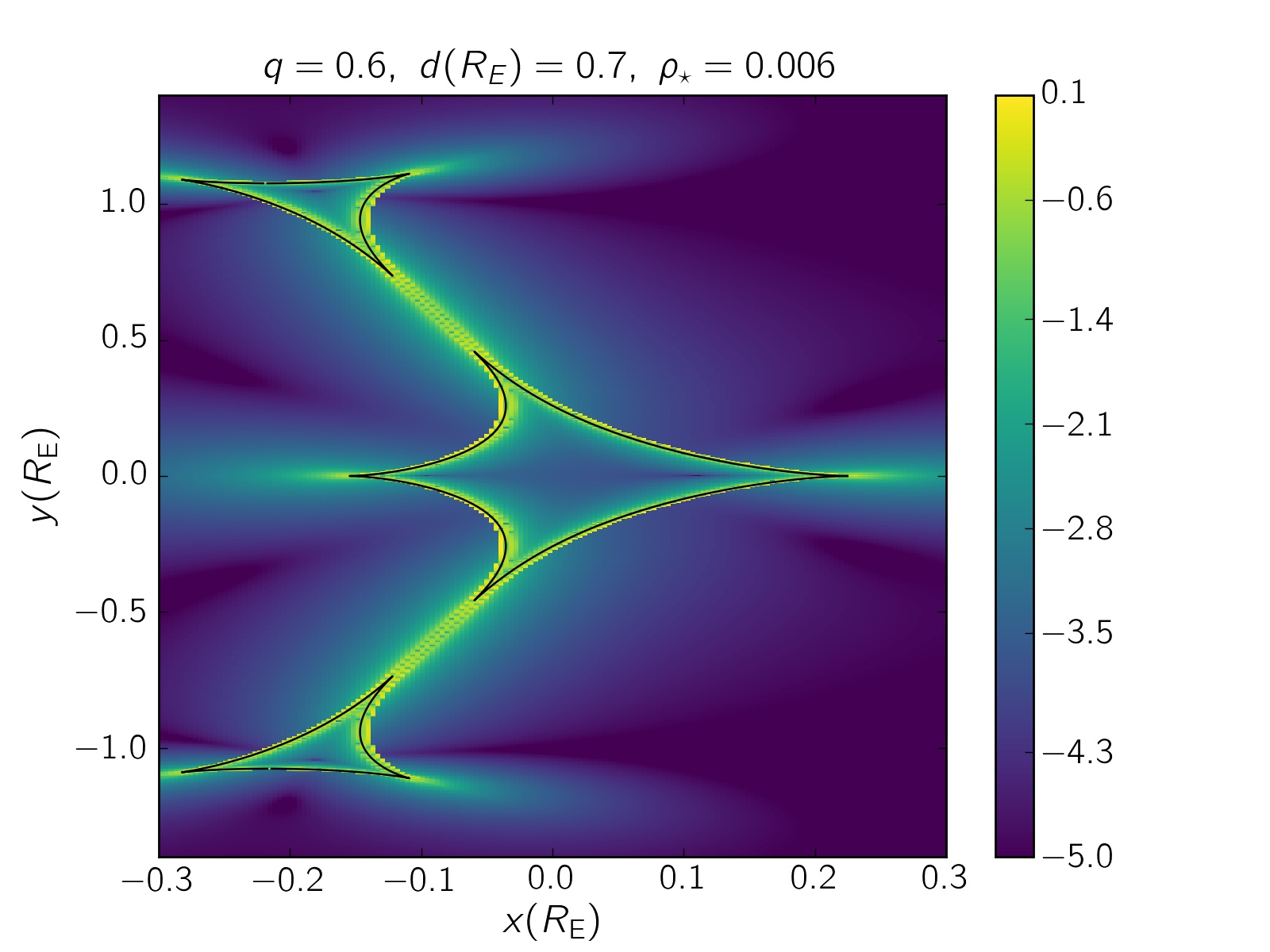}\label{fig1b}}
\subfigure[]{\includegraphics[angle=0,width=0.495\textwidth,clip=0]{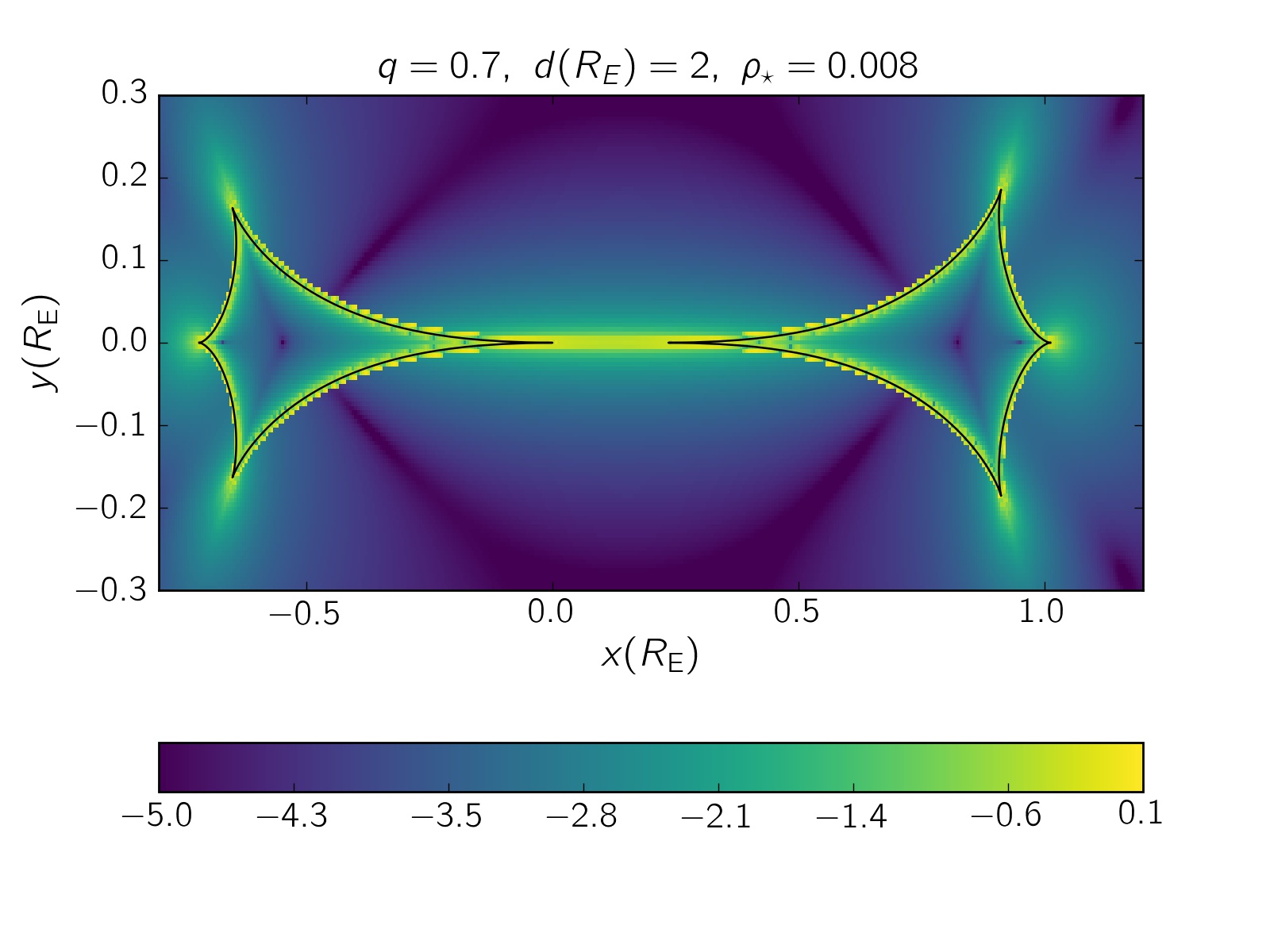}\label{fig1c}}
\subfigure[]{\includegraphics[angle=0,width=0.495\textwidth,clip=0]{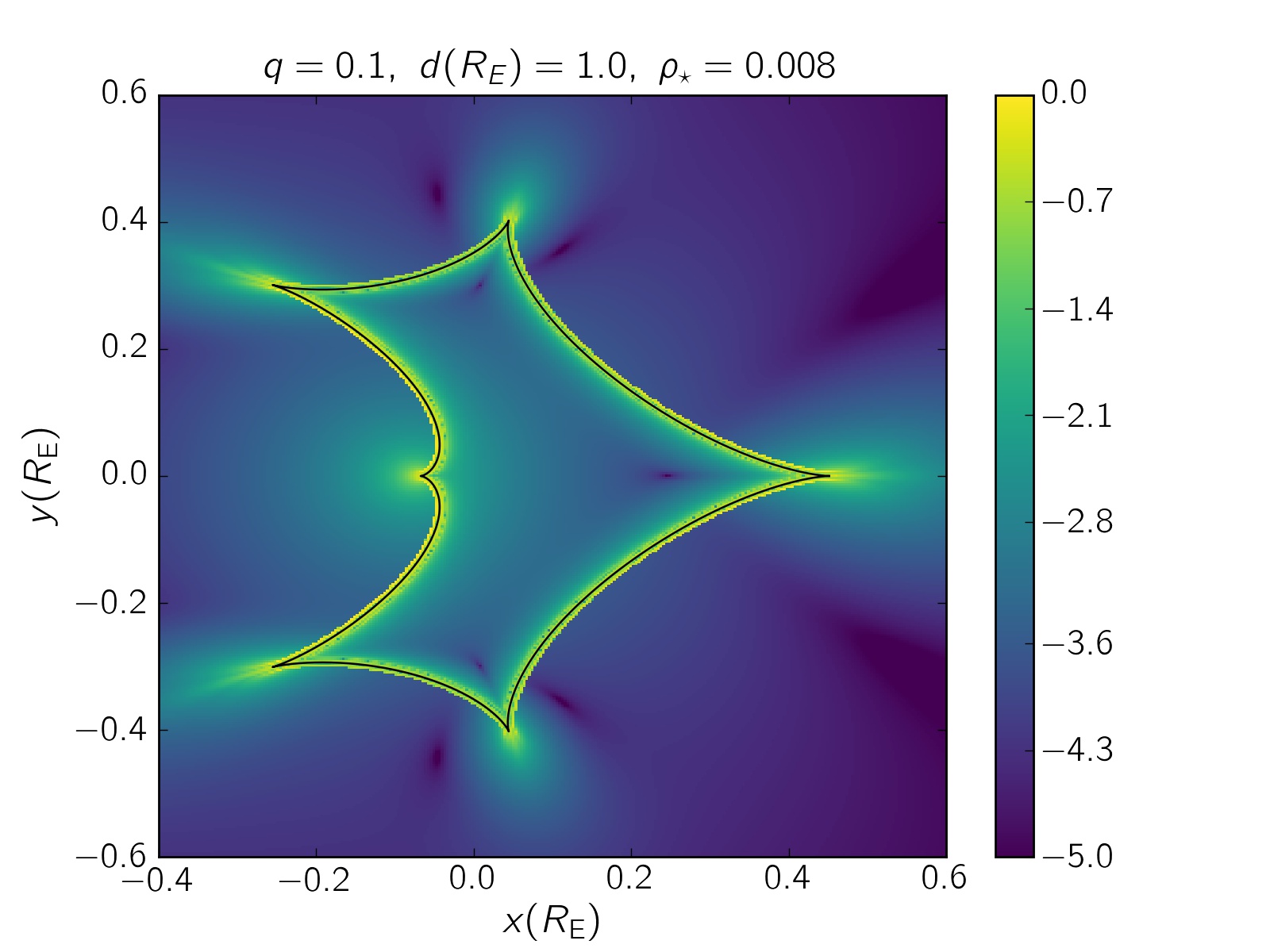}\label{figd}}
\caption{Polarization maps for four caustic configurations. The
parameters used to generate each of these maps, the lenses mass
ratio $q$, the microlenses separation normalized to the Einstein
radius $d(R_{\rm{E}})$ and the normalized source radius projected on
the lens plane $\rho_{\star}$ can be found at the top of each map.
The maps are in the logarithmic scale of polarization amounts in
percent. The black solid lines represent the corresponding caustic
curves.}\label{maps}
\end{figure*}
\section{Polarization in binary microlensing}\label{two}
The value of the polarization signal in a microlensing event depends
on (i) how much the circular symmetry of the source surface breaks,
(ii) the intrinsic and local polarization of the source star which
in turn depends on the scattering species and (iii) the source
radius. Throughout the paper, we consider hot and early-type stars
as sources. Noting that the different types of source stars have
similar patterns of the local polarizations. Hence, we expect
similar polarization behaviors but in different scales for different
stellar types. In early-type stars with a free electron atmosphere,
mostly Thomson scattering produces the polarization signal. Their
local Stokes (total and polarized) intensities \citep{Tinbergen1996}
were first analytically evaluated by \citet{chandrasekhar60} which
can be estimated as:
\begin{eqnarray}\label{II}
I_{I}(\rho)&=&I_{0}~[1-c_{1}(1-\mu)],\nonumber\\
I_{P}(\rho)&=&I_{0}~[c_{2}(1-\mu)],
\end{eqnarray}
where $\mu=\sqrt{1-\rho^{2}}$, $\rho$ is the radial position of each
point over the stellar surface normalized to the stellar radius and
projected on the sky plane and $I_{0}$ is the emergent radiative
flux of the source star from its center. $c_{1}=0.64$ and
$c_{2}=0.032$ are the so-called limb-darkening coefficients which
were evaluated by \citet{schneider1987}, based on the Chanrasekhar's
analytical solution of the radiative transfer equations in a
plane-parallel atmosphere. In this work, we use these amounts
throughout the paper.

These limb-darkening coefficients generally depends on the
wavelength, stellar surface temperature, its surface gravity and
partly the stellar metalicity. These dependencies can be determined
by numerically solving the radiative transfer equations in different
models of stellar atmospheres. For instance, \citet{Harrington15}
evaluated the monochromatic opacities and Stokes intensities for
early-type stars
\footnote{\url{https://www.astro.umd.edu/~jph/Stellar_Polarization.html}},
based on the NLTE code TLUSTY \citep{tlusty2,tlusty1}
\footnote{\url{http://nova.astro.umd.edu/index.html}}. According to
his results, we plot the limb-darkening coefficients, $c_{1}$
(dashed curves) and $c_{2}$ (solid curves), versus monochromatic
wavelength for four stellar models with different surface
temperatures and gravities in Figure (\ref{hotstar}). In these
models, the metal abundance is the same as solar one. Accordingly,
$c_{2}$ (unlike $c_{1}$) generally decreases in the larger
wavelengths. Also,  these values of limb-darkening coefficients are
a bit smaller than those estimated based on the analytical solution
of radiative transfer equations given by \citet{chandrasekhar60}.
\footnote{In order to estimate the limb-darkening coefficients in
the standard filters, we integrated of the Stokes intensities over
the bandwidths of Johnson-Cousins UBVRI photometric filters. In
$\rm{B}$-band, $c_{1}=0.4$ and $c_{2}=0.024$ which are close to the
analytical values, i.e., $c_{1}=0.64$ and $c_{2}=0.032$.}
\begin{figure*}
\centering
\subfigure[]{\includegraphics[angle=0,width=0.495\textwidth,clip=0]{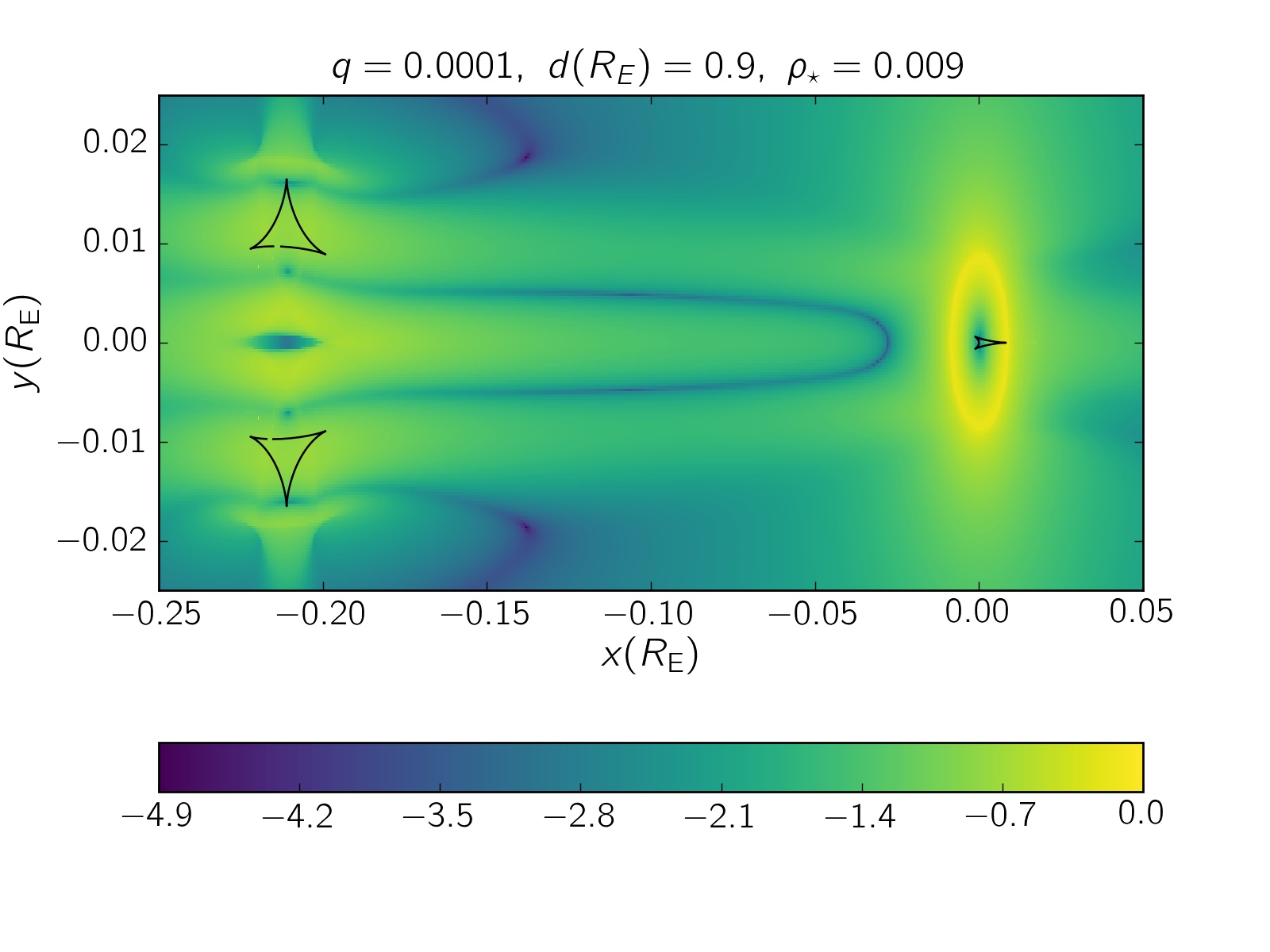}\label{figpa}}
\subfigure[]{\includegraphics[angle=0,width=0.495\textwidth,clip=0]{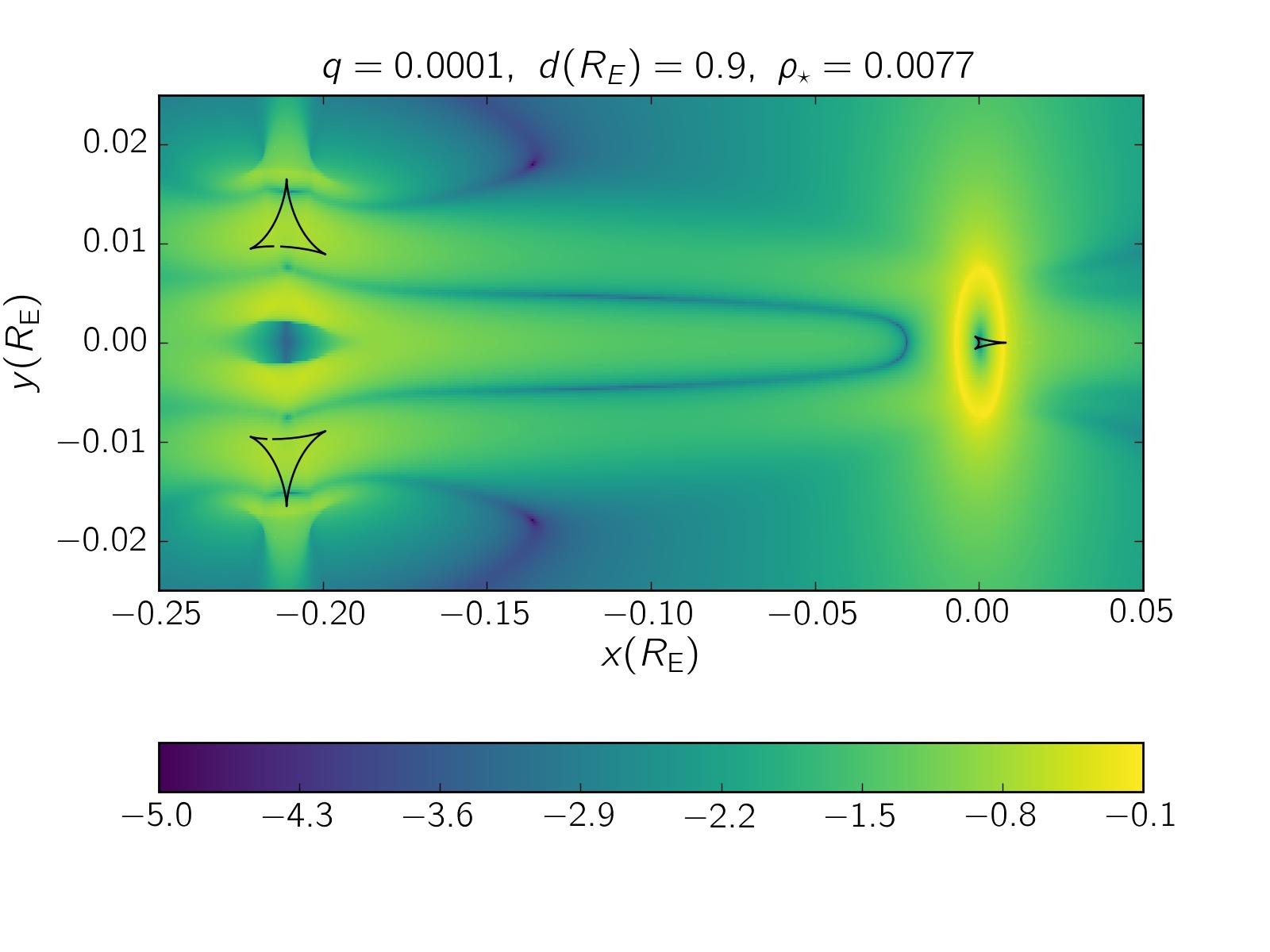}\label{figpb}}
\subfigure[]{\includegraphics[angle=0,width=0.495\textwidth,clip=0]{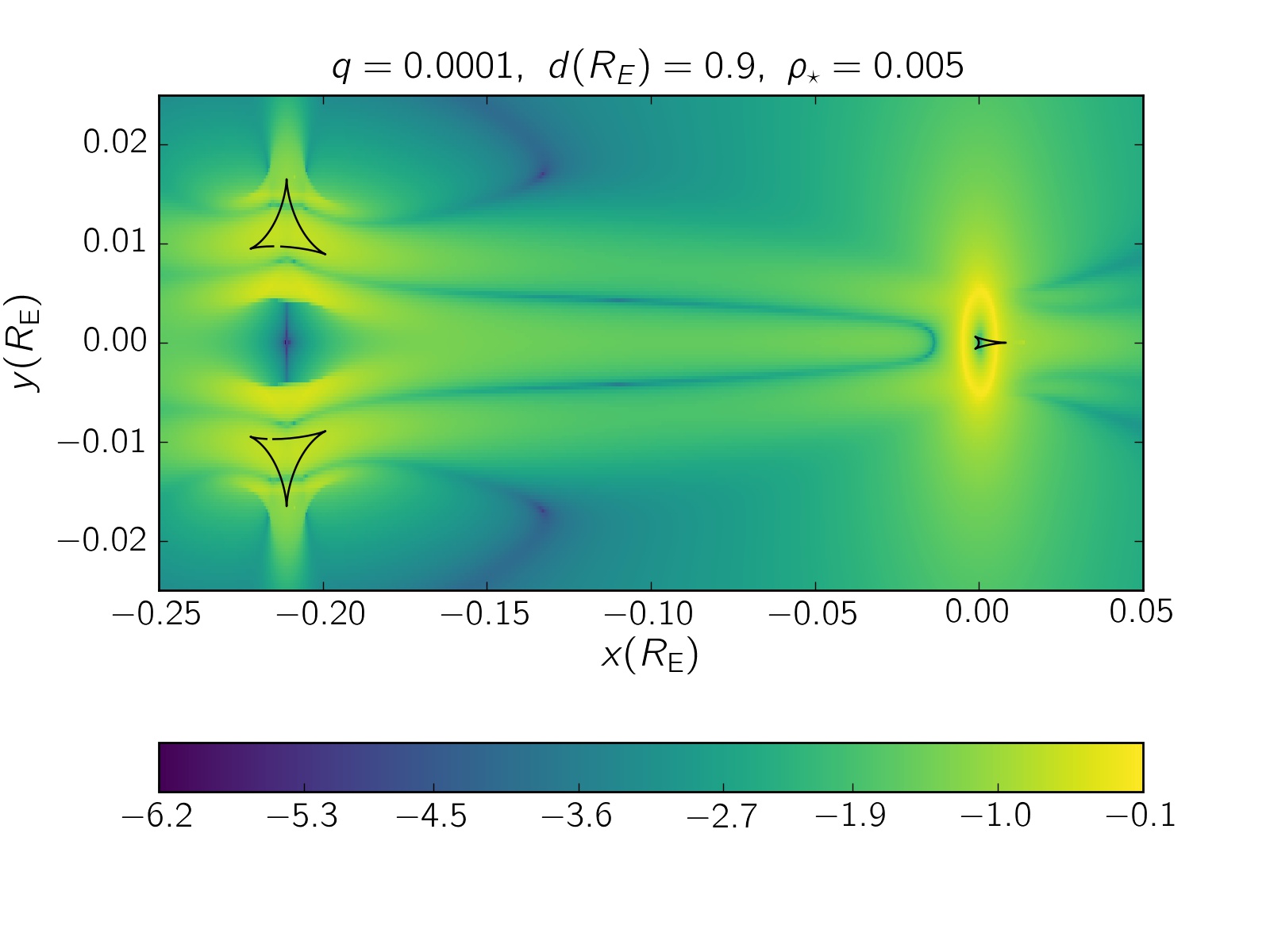}\label{figpc}}
\subfigure[]{\includegraphics[angle=0,width=0.495\textwidth,clip=0]{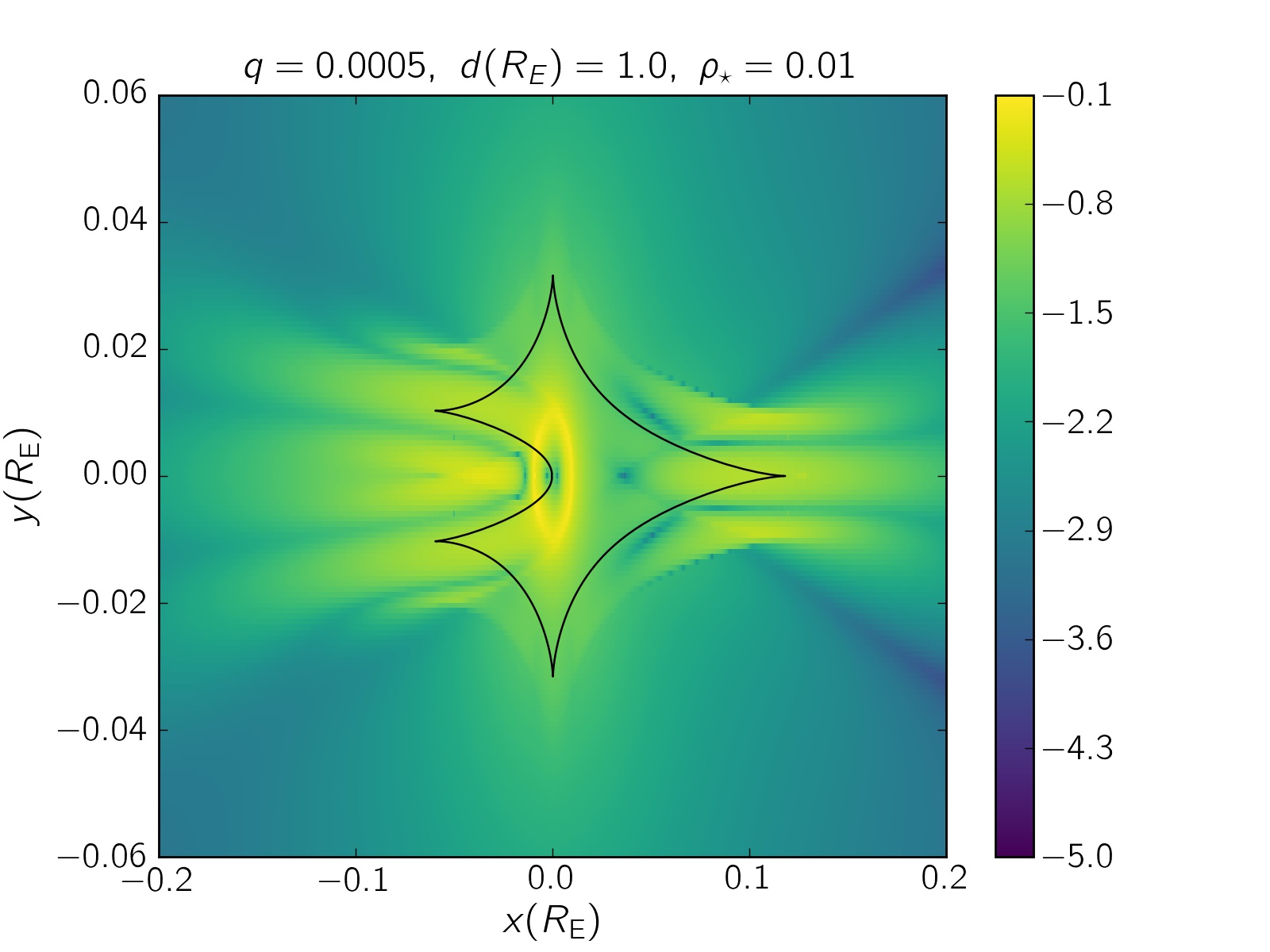}\label{figpd}}
\caption{Polarization maps for planetary caustic configurations. The
panels \ref{figpa}, \ref{figpb} and \ref{figpc} show the finite
source effects on the polarization maps of planetary caustics. In
these maps the source sizes are larger than the size of the central
caustics. In the last panel, the source size is on the order of the
size of the caustic. The maps are in the logarithmic scale of
polarization amounts in percent. The black solid lines represent the
corresponding caustic curves.}\label{planet}
\end{figure*}

In order to calculate the polarization signal during a microlensing
event, we integrate the local polarized and total Stokes intensities
(given by equations \ref{II}) over the star disk by considering the
magnification factor as a weight function. More details can be found
in many references, \citep[see, e.g.,][]{Ingrosso2012}. For
calculating the magnification factor in binary microlensing events
we use $\rm{RT}$-model which was developed by V.
Bozza\footnote{\url{http://www.fisica.unisa.it/GravitationAstrophysics/RTModel.htm}}
\citep{Bozza2018,Bozza2010,Skowron2012}. In this regard, we consider
each element of the source disk as a point-like source. The number
of elements depends on how much the source center
is close to the caustic line. 

Each binary-lens system as microlenses is specified by two factors:
$q$ the microlenses mass ratio and $d(R_{\rm{E}})$ the projected
separation between two microlenses normalized to the Einstein radius
$R_{\rm{E}}$. For a given binary microlenses and in every given
point in the lens plane, we calculate the polarization signal of the
source star while passing from that point which gives a map of
expected polarization signals. We call them as \emph{polarization
maps}. In the following subsection, we study the characteristics of
polarization maps for intermediate, close and wide topologies. Here,
we assume that the source size is very small with respect to the
size of the caustic.
\begin{figure}
\begin{center}
\includegraphics[angle=0,width=0.495\textwidth,clip=0]{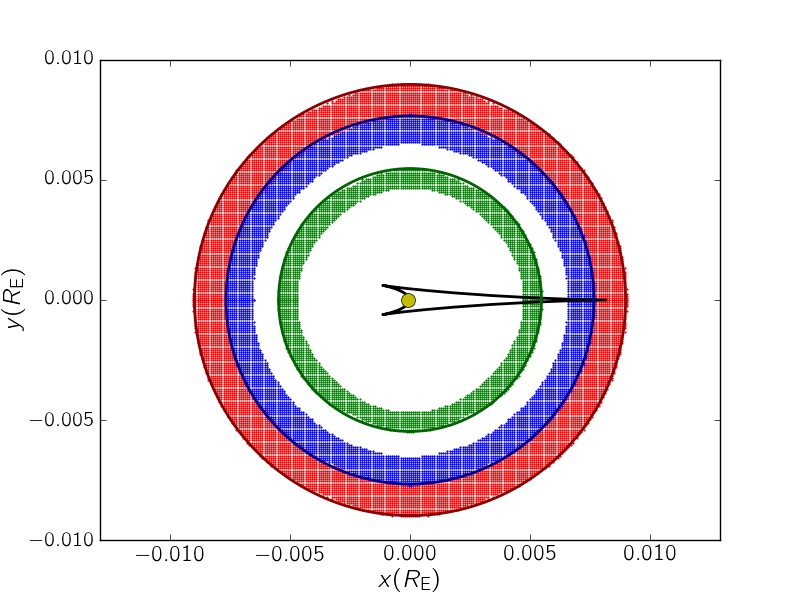}
\caption{The points on the lens plane with high polarization signals
$\geq 0.6\%$ around the central caustic of the planetary system
$(q=0.0001, d(R_{\rm{E}})=0.9)$ for three source radii
$\rho_{\star}=0.009$, $0.008$, $0.005$, represented with red, blue
and green filled points, respectively. The corresponding source
circles are shown by solid rings. The black curve and yellow point
are the central caustic and the position of the primary,
respectively.}\label{ring}
\end{center}
\end{figure}

\subsection{Polarization maps}\label{map}
In Figure (\ref{maps}), four examples of polarization maps for
different topologies are plotted. For better comparison with
magnification maps, we show the corresponding magnification maps in
Figure (\ref{ap1}) in Appendix. The polarization maps can be
described using some points which are classified in the following.

When the source radius is sufficiently smaller than the caustic
curve and in caustic crossings, the polarization signal maximizes
when the source edge is tangential to the caustic fold and its
center is out of the caustic curve. The local polarization signals
from the stellar edges are maximum (see equations \ref{II}). Hence,
whenever some part of the source edges magnifies highly, the
integrated polarization signal maximizes. Comparing different folds,
the maximum polarization signal occurs when the source is crossing
the folds create the closest cusp to the location of the primary,
although the highest magnification factor occurs when the source
crosses that cusp itself (see Figure \ref{ap1}). The value of these
maximum polarization signals depends on the source size, but mostly
reach to one percent for hot and early-type stars.
\begin{figure*}
\centering
\subfigure[]{\includegraphics[angle=0,width=0.495\textwidth,clip=0]{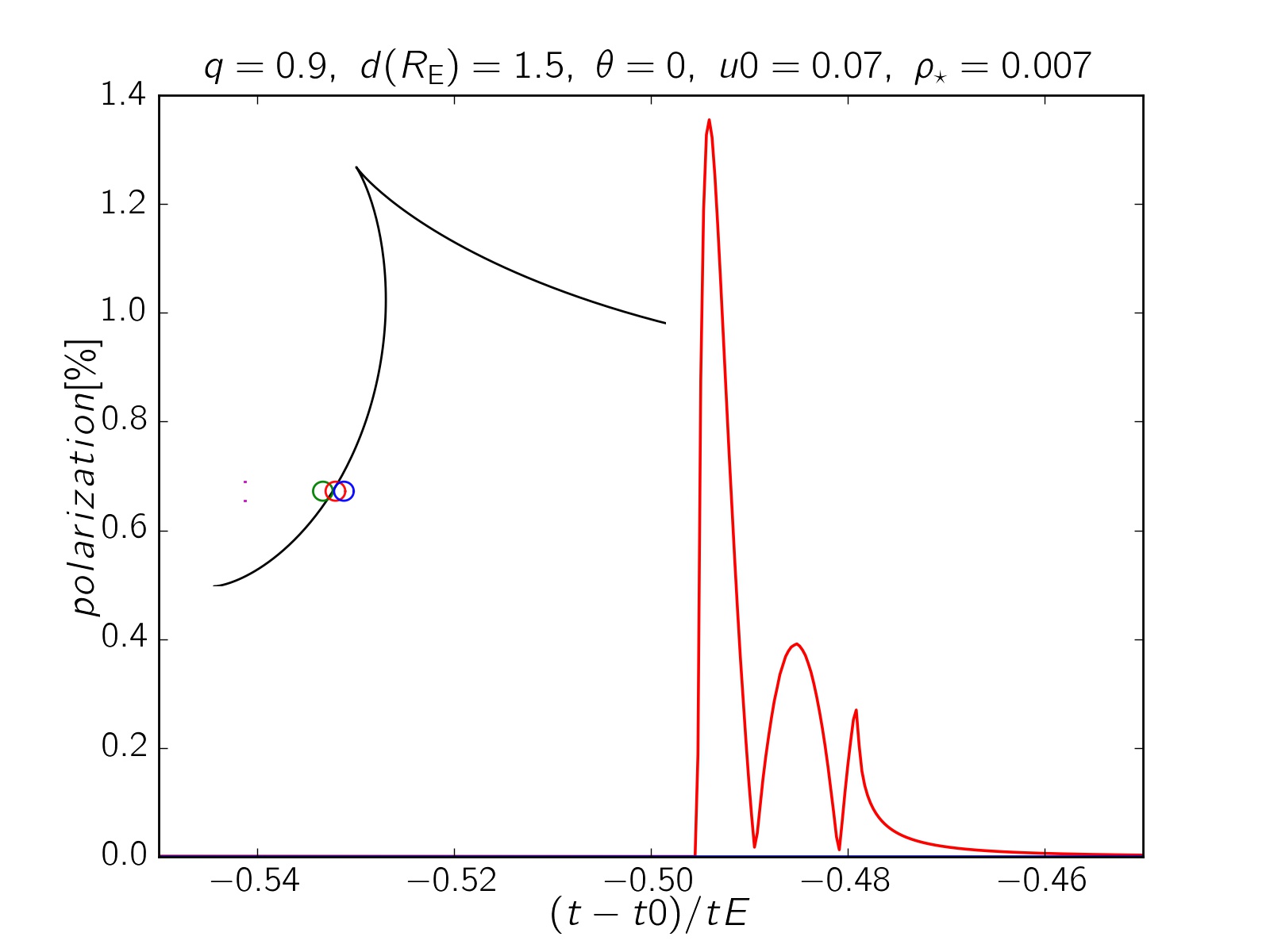}\label{fig2a}}
\subfigure[]{\includegraphics[angle=0,width=0.495\textwidth,clip=0]{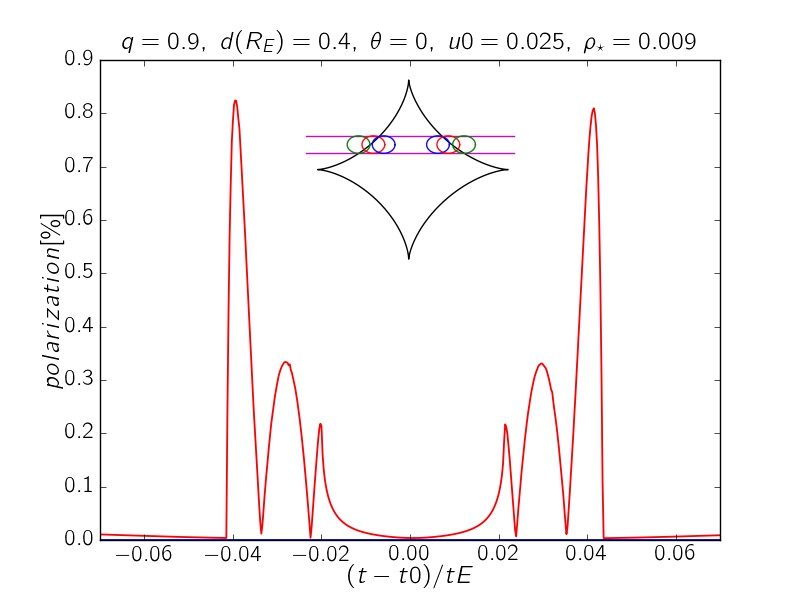}\label{fig2b}}
\subfigure[]{\includegraphics[angle=0,width=0.495\textwidth,clip=0]{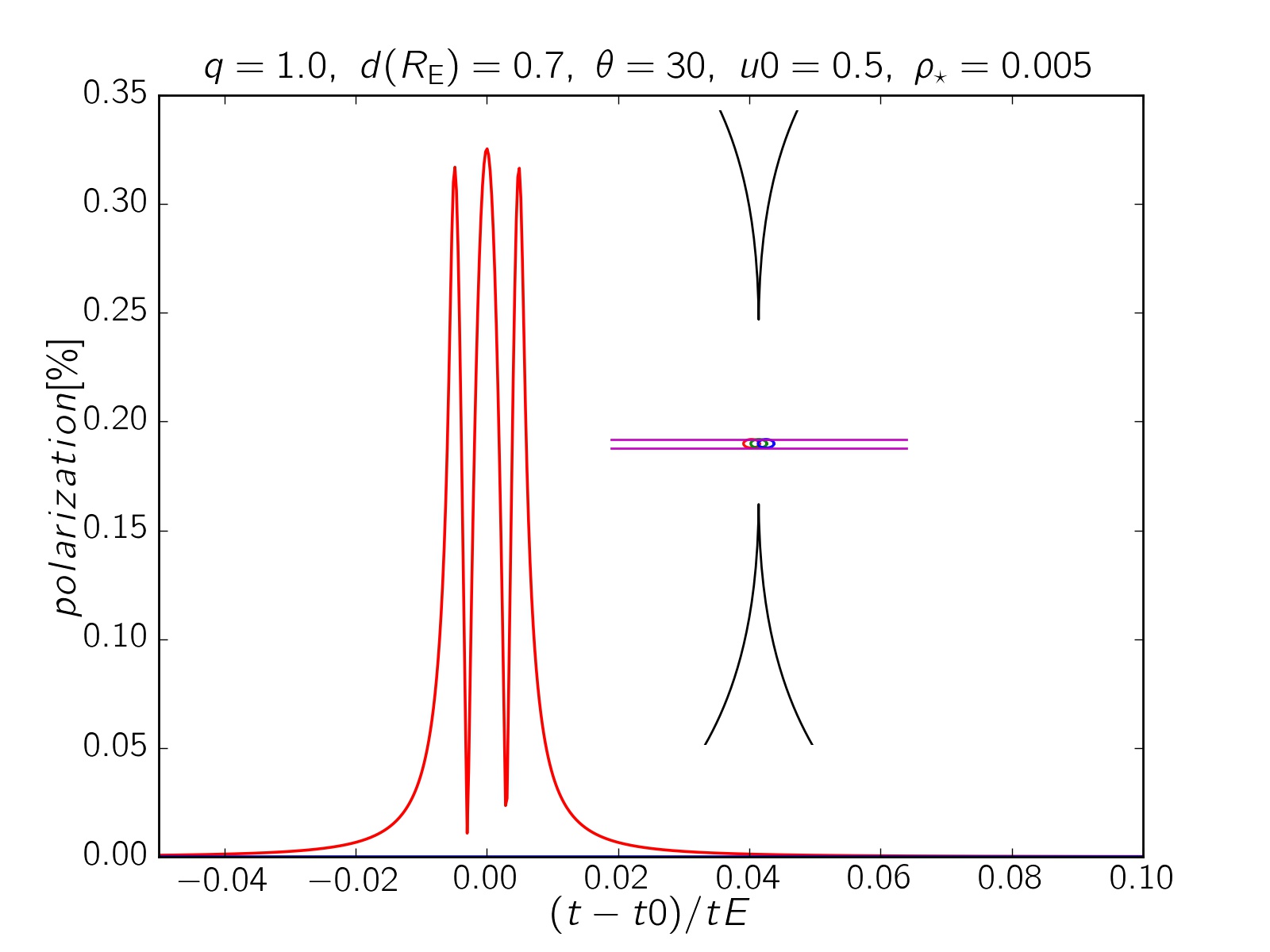}\label{fig2c}}
\subfigure[]{\includegraphics[angle=0,width=0.495\textwidth,clip=0]{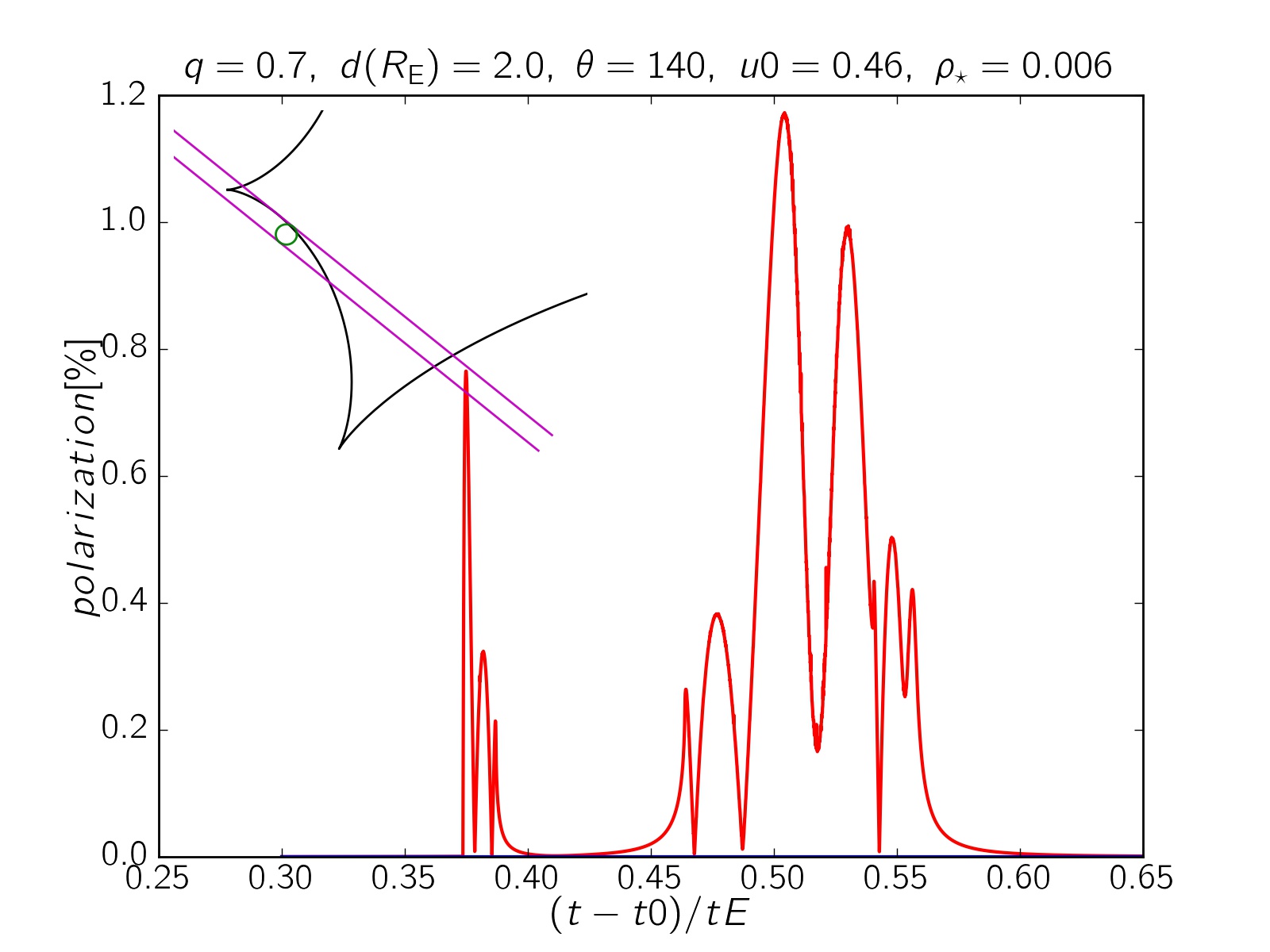}\label{fig2d}}
\caption{Four examples of the polarimetry curves while the source
star is passing from folds and the connection line between two
caustic curves. The insets show the trajectories of the source edges
on the lens plane (magenta lines), the fold curves (black lines) and
the source positions at the times of the highest (green circles),
middle (red circles) and smallest peaks (blue circles). The
parameters used to generate each curve are given at the top of each
plot. $\theta$ is the angle between the source trajectory and the
binary axes, given in degree.}\label{foldss}
\end{figure*}

Since the maximum local polarized and total intensities from the
source surface are from the stellar edge and its center,
respectively, so the magnification peak takes place when the source
center is on the fold and its the time interval with the largest
polarimetry peak is $t_{p}\simeq0.96~t_{\star}$, where
$t_{\star}=t_{\rm{E}}~\rho_{\star}$ is the time of crossing the
source radius. When entering the caustic line, the polarization peak
happens earlier whereas in exiting from the caustic line the
magnification peak happens earlier. When the source does not cross
the caustic and just passes close to it, their time interval depends
on the caustic shape, the source trajectory and the source radius.
This time interval decreases by increasing the impact parameter.

When the source is wholly inside the caustic curve, although its
magnification factor is considerable and it is not zero, the
polarization signal is almost zero. In that case, every point over
the source disk is magnified, but there is no significant symmetry
breaking.

On the connection lines between different caustic curves in close
and wide topologies, the polarization signals are considerable,
although they do not reach to one percent, which are achievable in
fold caustic crossings (see, e.g., Figure \ref{fig2c}).

Comparing the magnification and polarization maps, whenever the
magnification map is smooth and without high fluctuations, there is
no significant polarization signal, even though the magnification
factor is high at every point, for instance inside the caustic
curve.

\subsection{Polarization in planetary microlensing}\label{size}
When microlenses constitute a planetary system, we have one small
central and two planetary caustics. For very small mass ratio, the
size of the central caustic is proportional to $\propto q$
\citep{chung2005} and mostly small in comparison with the source
size. We plot four polarization maps corresponding to some
planetary systems 
in Figure (\ref{planet}). Their corresponding magnification maps are
shown in Figure (\ref{ap2}) of Appendix. According to these maps, in
planetary microlensing events the maximum polarization signals take
place over a ring around the position of the primary. Noting that
the magnification factor maximizes at the center of this ring (see
Figure \ref{ap2}). Hence, whenever the source edge passes from the
location of the primary its polarization signal maximizes.

\begin{figure*}
\centering
\subfigure[]{\includegraphics[angle=0,width=0.495\textwidth,clip=0]{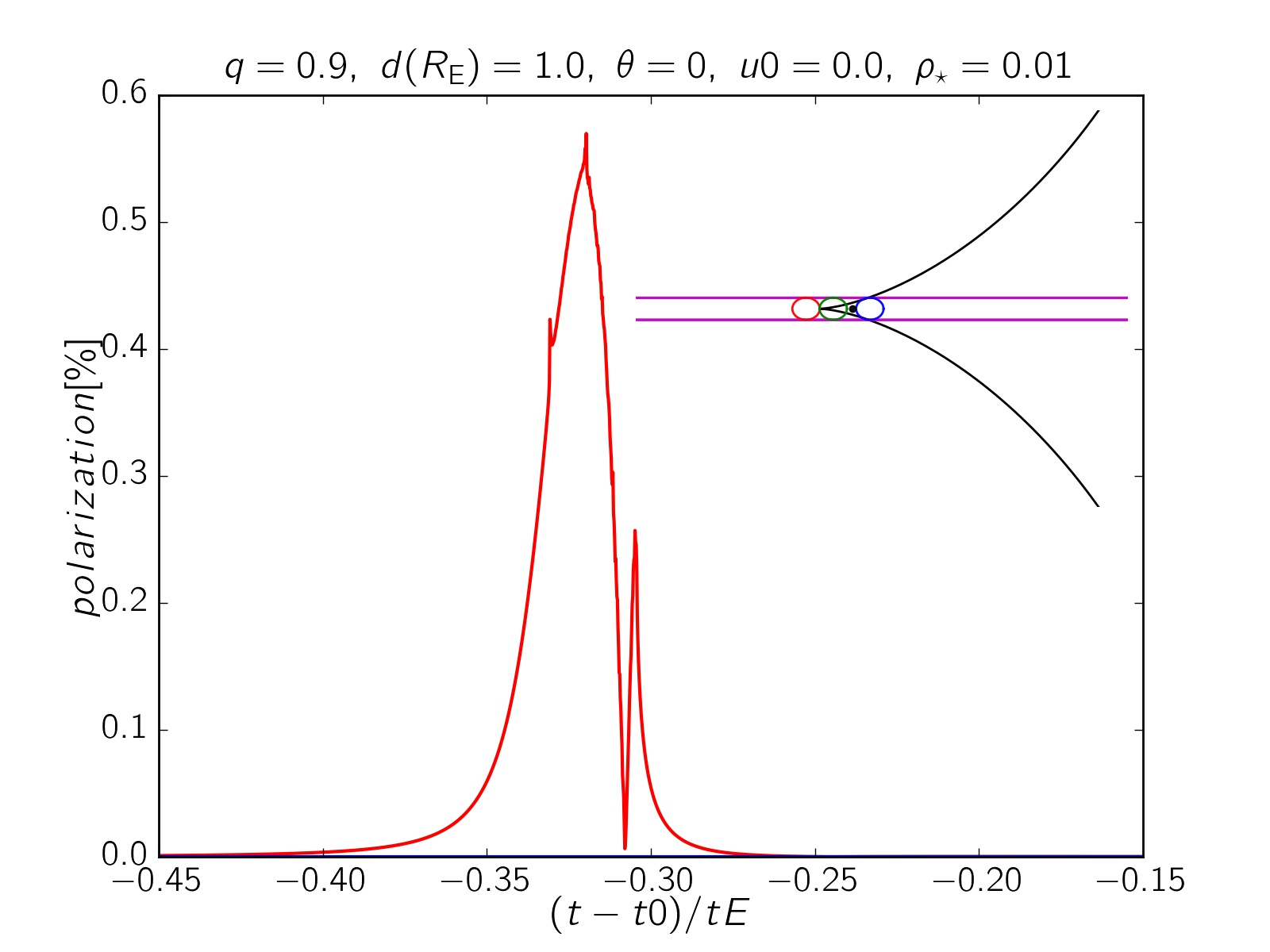}\label{fig3a}}
\subfigure[]{\includegraphics[angle=0,width=0.495\textwidth,clip=0]{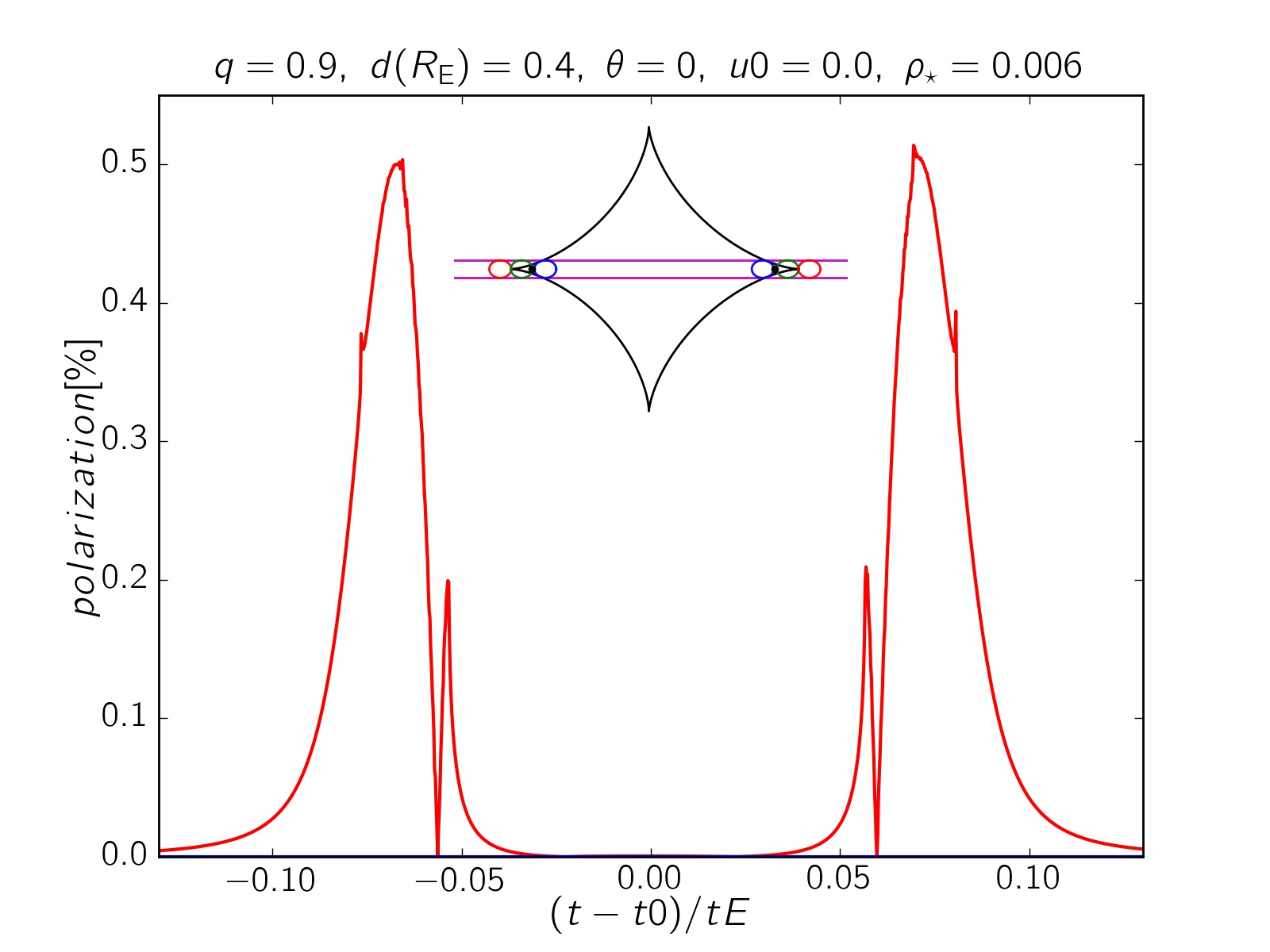}\label{fig3b}}
\subfigure[]{\includegraphics[angle=0,width=0.495\textwidth,clip=0]{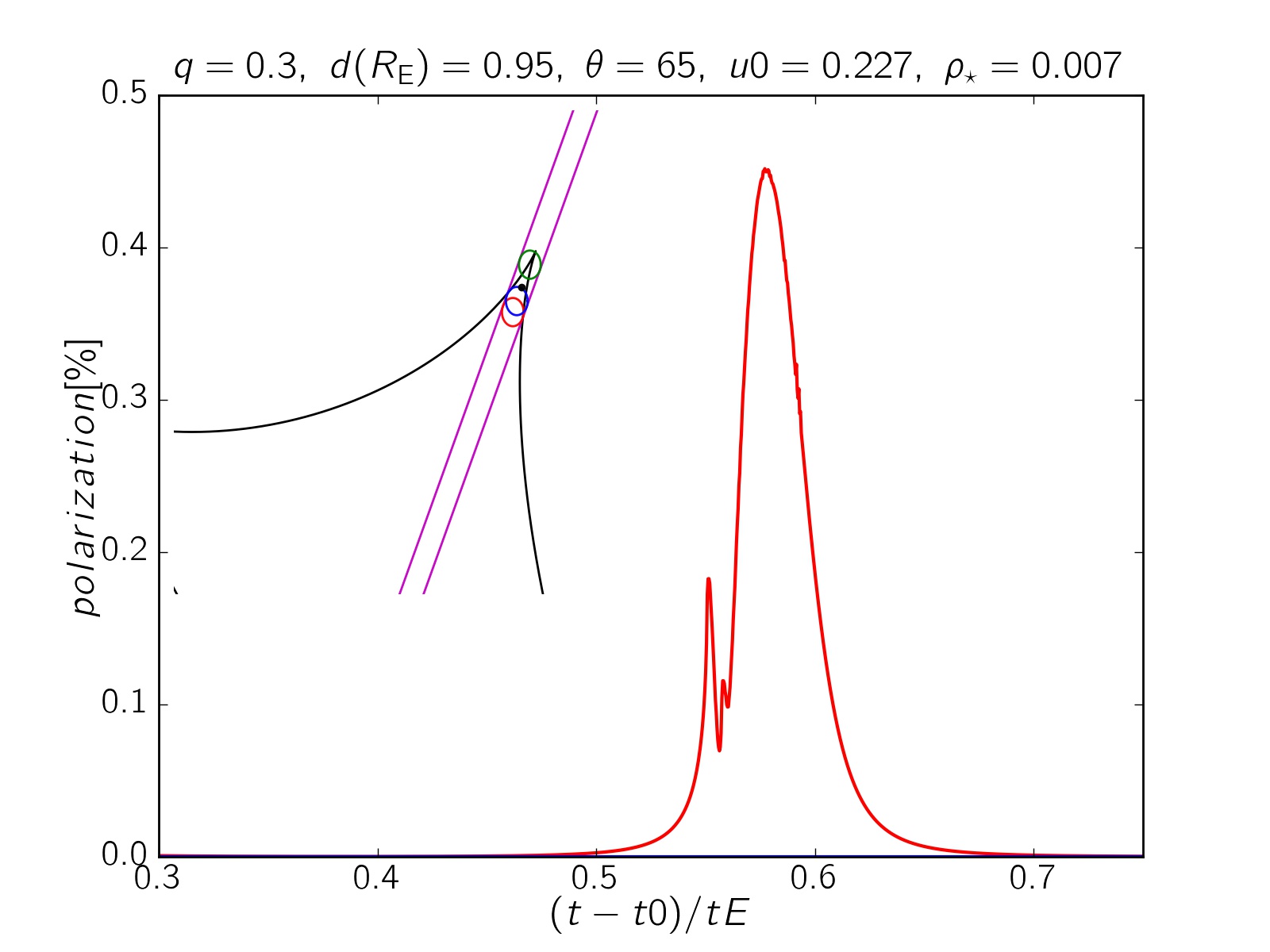}\label{fig3c}}
\subfigure[]{\includegraphics[angle=0,width=0.495\textwidth,clip=0]{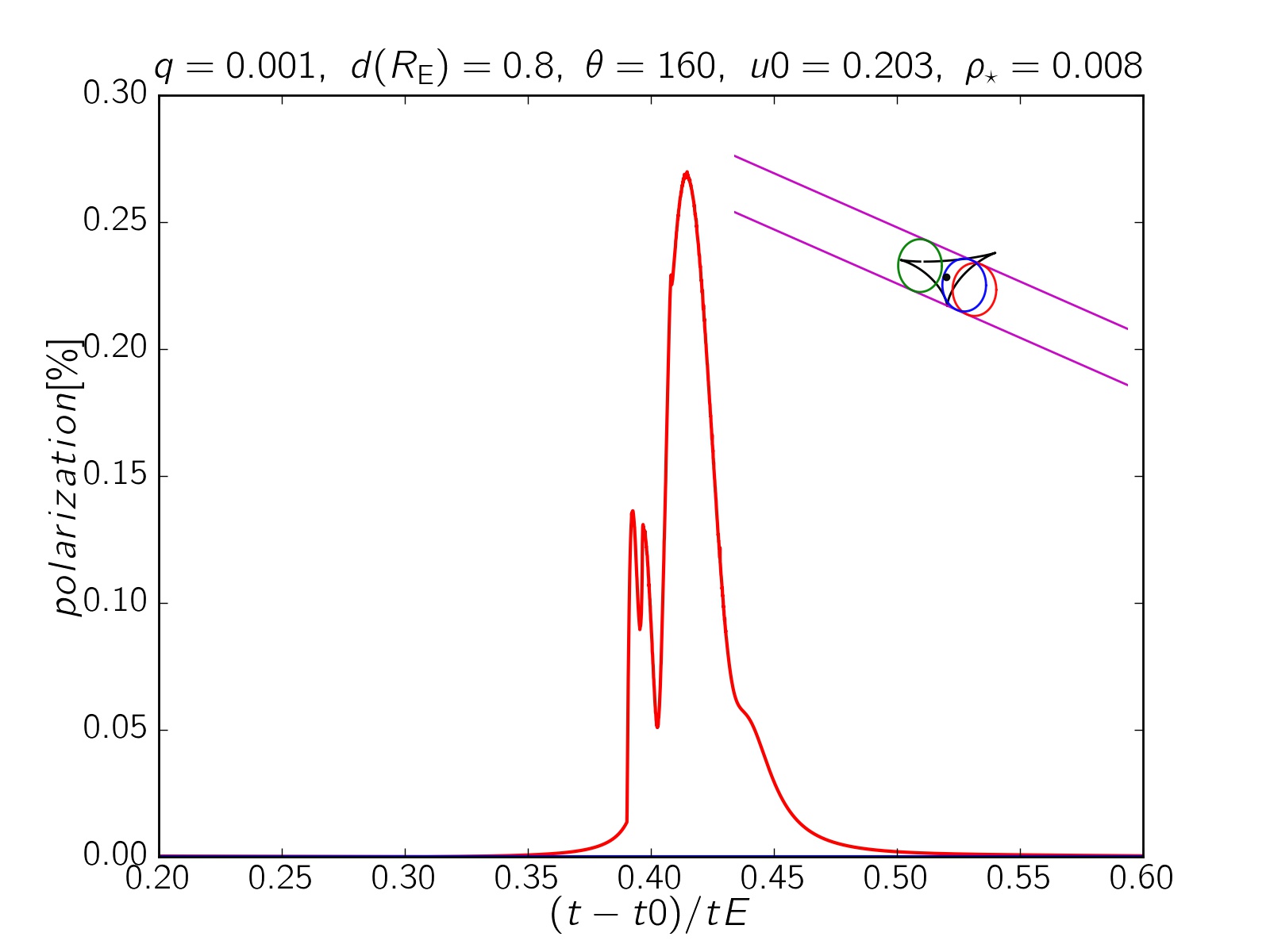}\label{fig3d}}
\caption{Four examples of the polarization curves in binary
microlensing events and during cusp crossing. More details can be
found in the caption of Figure (\ref{foldss}).}\label{cusp}
\end{figure*}
In order to study the finite source effect on the polarization maps
in planetary microlensing events, in Figures \ref{figpa},
\ref{figpb} and \ref{figpc} we consider three different source
radii, i.e., $\rho_{\star}=0.009,~0.008,~0.005$. The larger source
radius, the greater ring with higher polarization amounts. According
to the numerical calculations, the radius of the rings is
$\sim0.96\rho_{\star}$. Hence, the time interval between the
polarization and magnification peaks is $0.96~t_{\star}$, when the
source passes from the central caustic.

To better show the \emph{circular} shape of these rings and compare
their radii with the source radius, the rings are plotted in Figure
(\ref{ring}). In this plot, the points with the polarization signals
higher than $0.6$ percent (most likely measurable) are shown. The
red, blue, green points and circles are the high polarization points
and the source rings corresponding to
$\rho_{\star}=0.009,~0.008,~0.005$, respectively. The yellow circle
represents the position of the primary. The central caustic is
plotted with the black curve.

The magnification factor is not high around planetary caustics (see
Figure \ref{ap2}), whereas the polarization signals is not zero
around them. However, their polarization signals can not reach to
one percent and is less than those upon the rings around the central
caustic. Right between two parts of planetary caustics the
polarization is very low, because of symmetry.

The last map of Figure (\ref{planet}), is an intermediate topology
with small value for $q$. The size of the resonance caustic is
$\propto q^{1/3}$ \citep{gaudi2012} for a fixed value of $d$. We
choose the source size as $\rho_{\star}=0.01$ on the same order of
magnitude of the size of the caustic for
$q=0.0005,~d(R_{\rm{E}})=1.0$. Similarly, the highest polarization
signals take place on a circular ring whose center is at the
location of the perimetry. If the source passes normal to the binary
axis (when it is outside the caustic), three polarization peaks
create. The polarization angle at the position of each peak differs
by $90^{\circ}$ with respect to its next peak.

The photometry and astrometry behaviors of source stars while fold
and cusp caustic crossings were studied in many references well
\citep[see, e.g.,][]{gaudi2002b,dominik2004,gaudi2002}. In the
following subsections, we study the polarization curves in fold and
cusp crossings.

\subsection{Polarization in fold caustic crossings}\label{fold}
The polarization curve of a star while entering into (extracting
from) the fold of a caustic has three peaks, which happens (i) when
the first edge of the source is on the caustic line, (ii) when the
source center is on the caustic line and (iii) when the last edge of
the source star is on the caustic line. The first is the highest,
because a small part of the source edge is highly magnified and
other parts are not magnified much, which makes the largest
asymmetry. The last peak is the smallest, because all of the source
points magnify. The middle one is the widest and occurs at the same
time of the magnification peak. The time interval between two
consecutive peaks is $t_{\rm{p}}$. Hence, the polarization curve
during fold caustic crossings lasts about $2~t_{\rm{p}}$. At the
time of the middle peak, the polarization angle differs from that at
the time of other peaks by $90^{\circ}$. Figure (\ref{foldss}) shows
four polarization curves of fold-crossing microlensing events. The
green, red and blue circles show the source rings at the time of the
highest, widest and smallest polarization peaks.

In the panel \ref{fig2c}, the source star is traveling from the
connection line between two caustic curves. In that case we have
three polarization peaks. The middle one occurs at the same time of
the magnification peak. Their size and time intervals depend on the
angle between the source trajectory and the connection line. For
symmetric cases, i.e., when the source trajectory is normal to the
connection line, (i) the middle peak is the highest and two side
peaks have symmetric shapes and at the same time intervals with
respect to the middle. (ii) The polarization angle from the first
peak to the second peak changes by $90^{\circ}$ and from the second
one to the last one changes by $90^{\circ}$.

\subsection{Polarization in cusp caustic crossings}\label{cusps}
While cusp caustic crossings all polarization curves have similar
behaviors. In Figure (\ref{cusp}), we show four examples of
polarization curves for binary-lens microlensing events with cusp
caustic crossings. In the panels \ref{fig3a} and \ref{fig3b} the
source trajectories pass on the symmetric axis of cusps (which
divides the cusps into two symmetric parts) and in the two others
panels the source trajectories do not coincide on the symmetric axes
of cusps.

Two first polarization curves have the same behaviors. They have
three peaks. The first peak is very small and occurs at the time
when the source edge is on the corner of the cusp and its center is
out of the caustic (the source ring at this time is shown as red
circle). The second peak is the largest and widest one (main peak)
and takes place when the source edge is on the corner of the cusp
while its center is inside the caustic (shown as green circles). The
polarization angles at the times of these peaks are the same and
their time intervals are exactly $2t_{\star}$ (green and red circles
are tangential). The last one happens when the source is completely
inside the caustic curve and its last edges are tangential to the
folds which meet at that cusp (shown as blue circles). The
polarization angle at the time of this peak differs by $90^{\circ}$
from two previous peaks.

The magnification peak in cusp caustic crossings happens at some
time between the main and third peaks. The source positions at the
time of the maximum magnification signals are shown as filled black
circles which is upon the edges of blue circles. Hence, the time
interval between the magnification and third polarization peak is
$t_{\star}$. While entering the cusp the magnification peak happens
earlier and while existing from the cusp that polarization peak
occurs earlier.

In the panel \ref{fig3c}, the source trajectory does not coincide on
the symmetric axis of the cusp and its polarization curve is
somewhat different from ones shown in panels \ref{fig3a} and
\ref{fig3b}. According to Figure \ref{fig3c}, in this asymmetric
case the polarization curve has three peaks. The largest and widest
one happens when the source edge is on the corner of the cusp and
the source center is into the caustic curve, the source at the time
of this peak is shown as green circle. This peak is similar to the
main peak in symmetric cases. When the source is completely into the
caustic curve two other peaks appear: One of them at the time that
the source edges are tangential to the folds meet at the cusp (the
source ring is shown as a blue circle) and the other at the time of
the last connection of the source edge and the folds (shown with red
circle). The time of the last peak and its value depends on the
source trajectory, its radius and the cusp shape.

The polarization angles at the time of peaks (i) and (ii) differ by
$90^{\circ}$. The magnification peak happens with the time interval
$t_{\star}$ before (after) the time of the middle peak while
entering into (existing from) the cusp.

In the panel \ref{fig3d}, the polarization curve when the source
size is almost equal to the planetary caustic and the source passes
from one of the caustic cusps. Its polarization curve has similar
behavior with the polarization curve plotted in the panel
\ref{fig3c}.

The significant point regarding the polarization curves while cusp
caustic crossings is that the polarization signal even at the
largest peaks do not reach to one percent for early-type stars,
whereas it reaches to one percent or even more while crossing the
folds join at the primary's location. In Figure \ref{fig2d} the
source star passes from both fold and cusp of a caustic curve. The
highest polarization signal occurs when the source is tangential to
the fold for a while. The source star is shown as a green circle at
the time of the maximum polarization in this plot.

While fold caustic crossings, the polarization reaches to zero
between two consecutive peaks, whereas in the cusp caustic
crossings, the polarization between peaks mostly does not achieve to
zero.

\begin{figure*}
\centering
\subfigure[]{\includegraphics[angle=0,width=0.495\textwidth,clip=0]{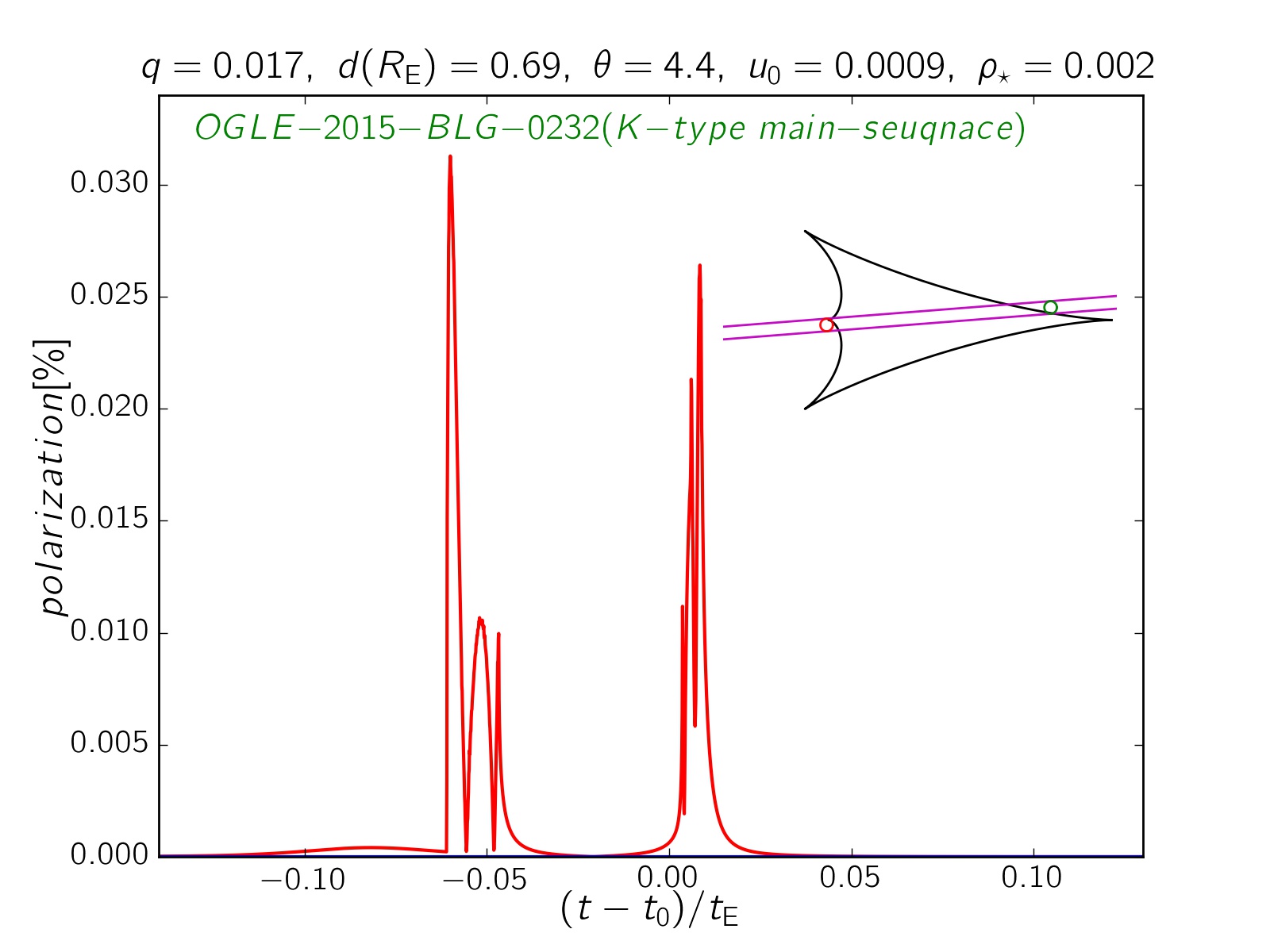}\label{fig4a}}
\subfigure[]{\includegraphics[angle=0,width=0.495\textwidth,clip=0]{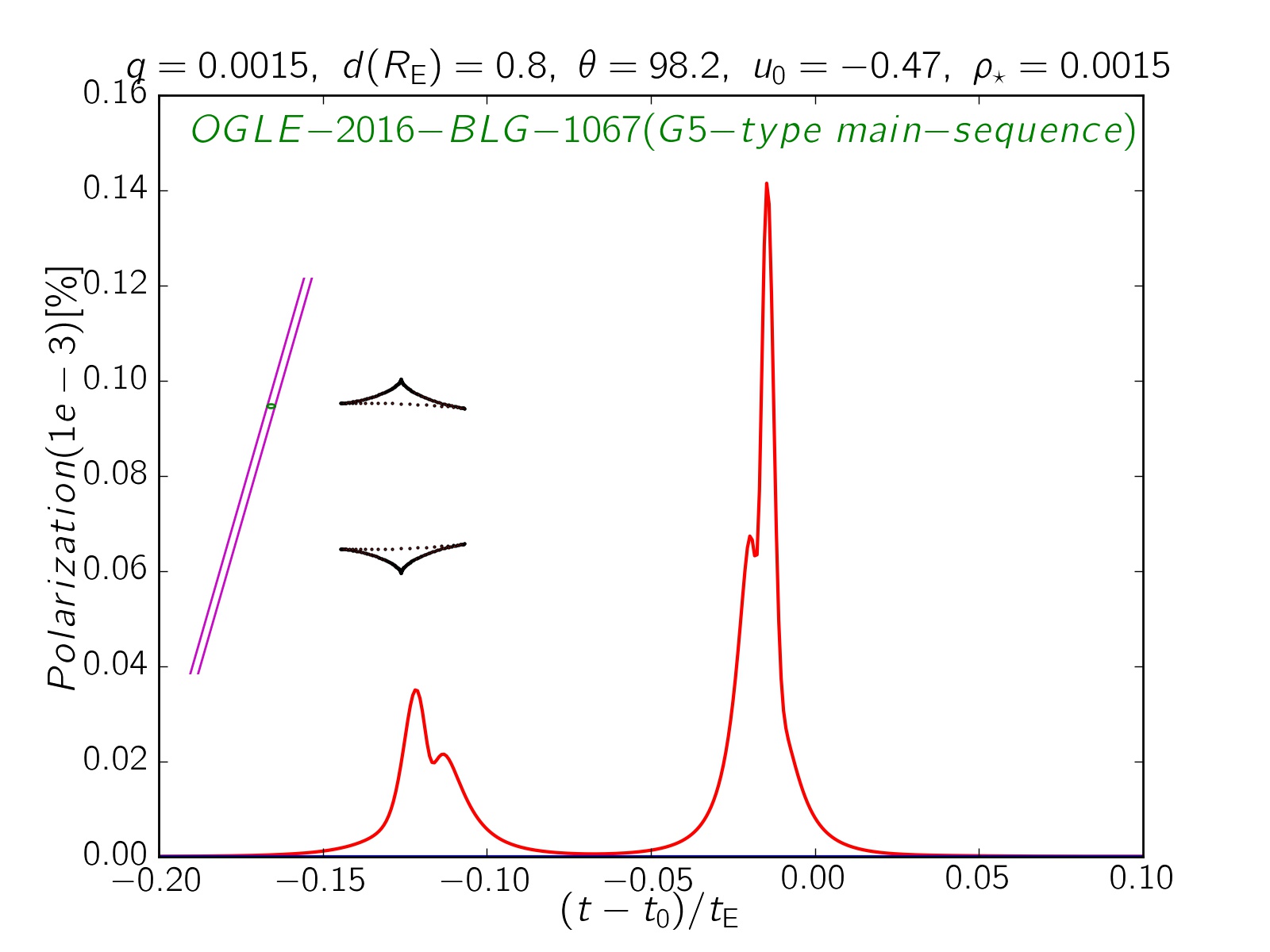}\label{fig4b}}
\subfigure[]{\includegraphics[angle=0,width=0.495\textwidth,clip=0]{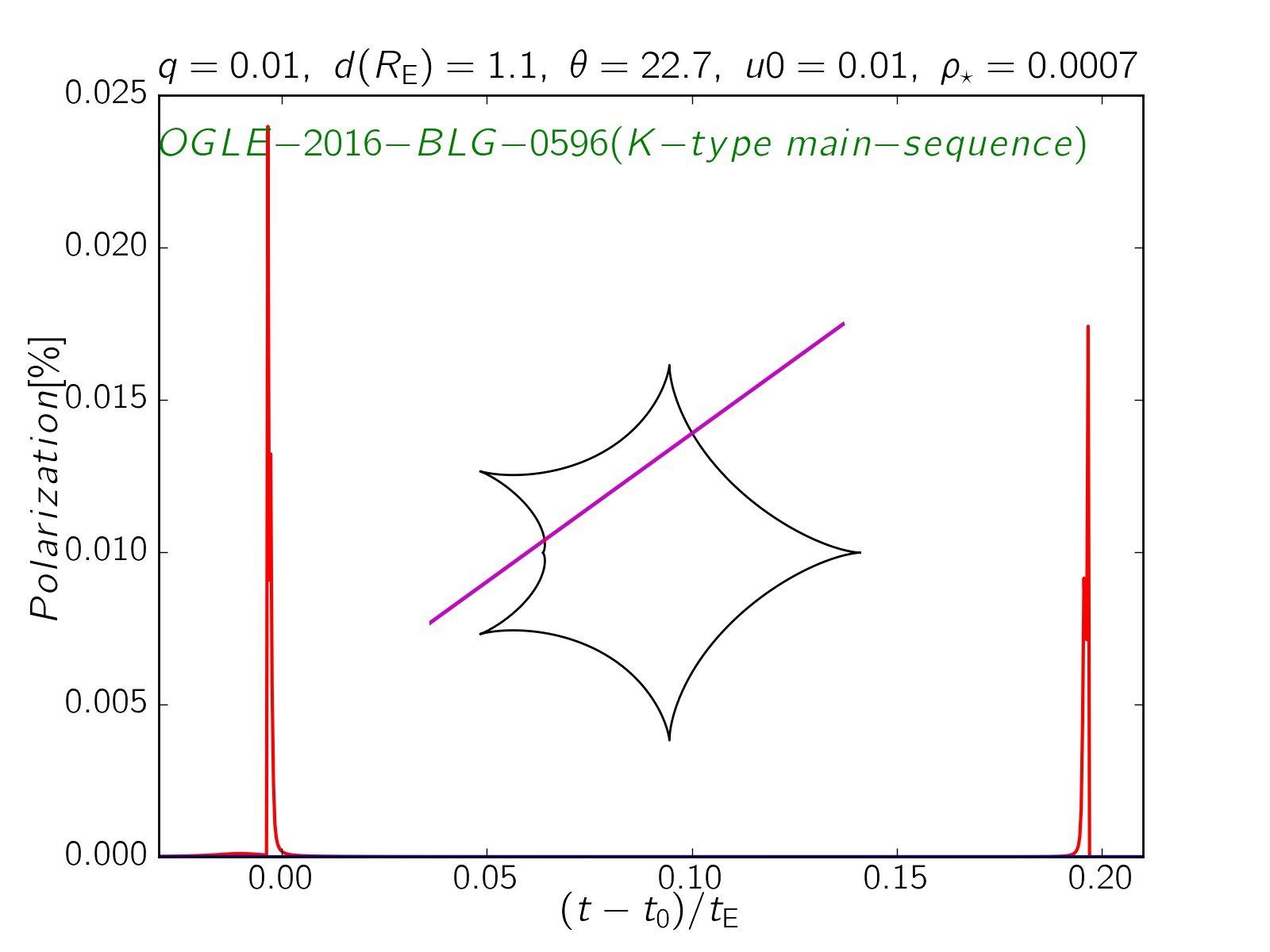}\label{fig4c}}
\subfigure[]{\includegraphics[angle=0,width=0.495\textwidth,clip=0]{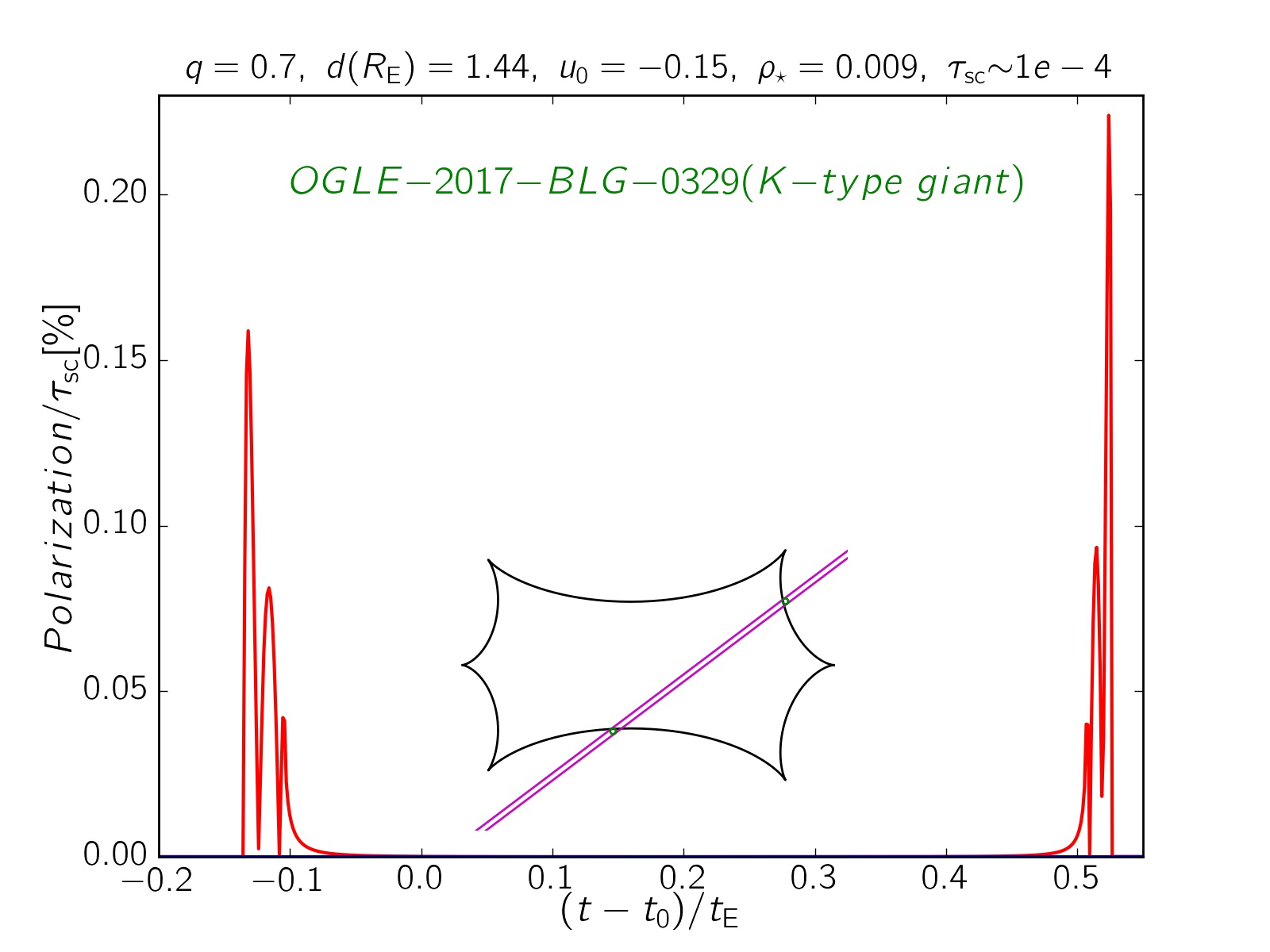}\label{fig4d}}
\subfigure[]{\includegraphics[angle=0,width=0.495\textwidth,clip=0]{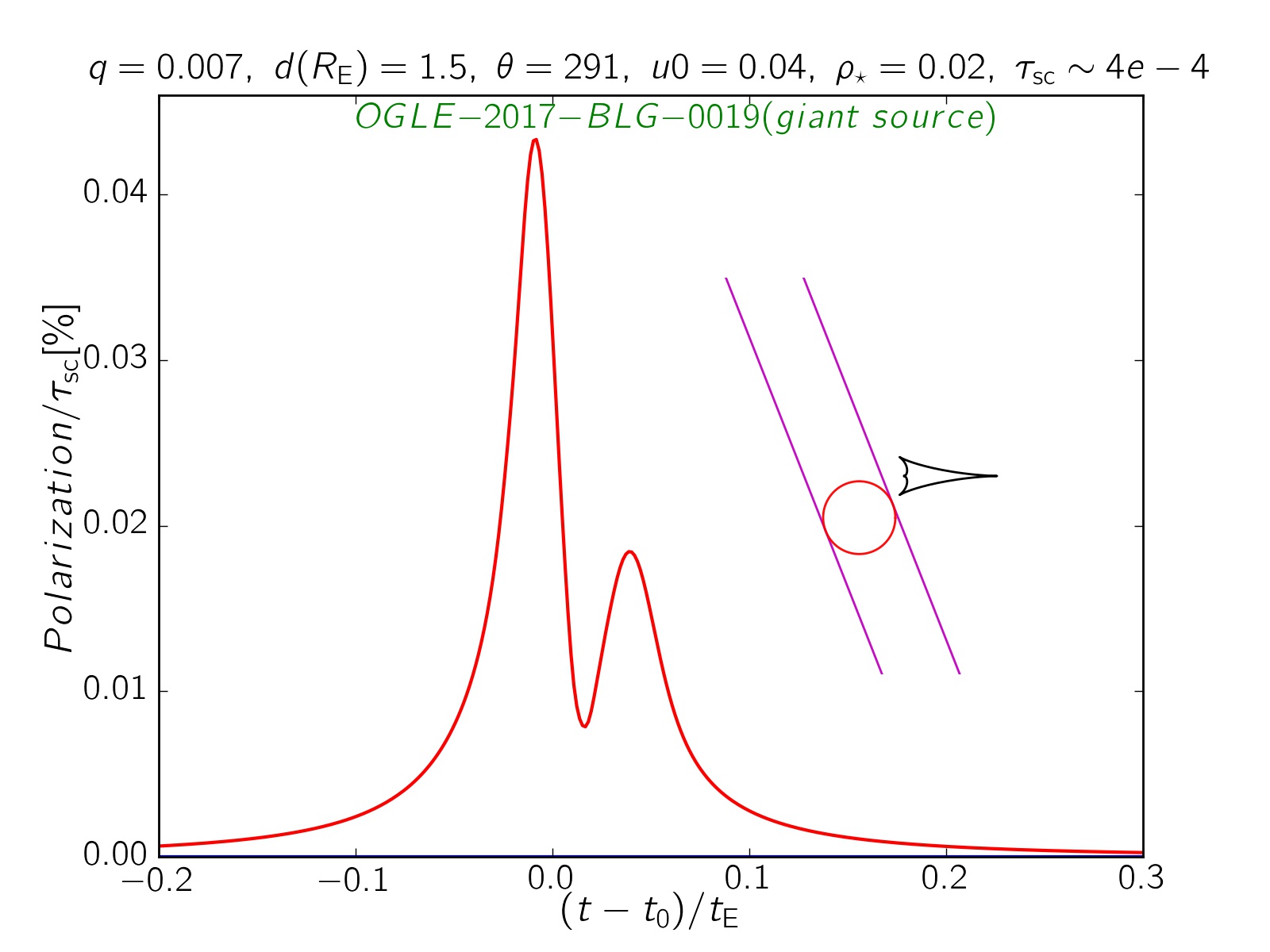}\label{fig4e}}
\subfigure[]{\includegraphics[angle=0,width=0.495\textwidth,clip=0]{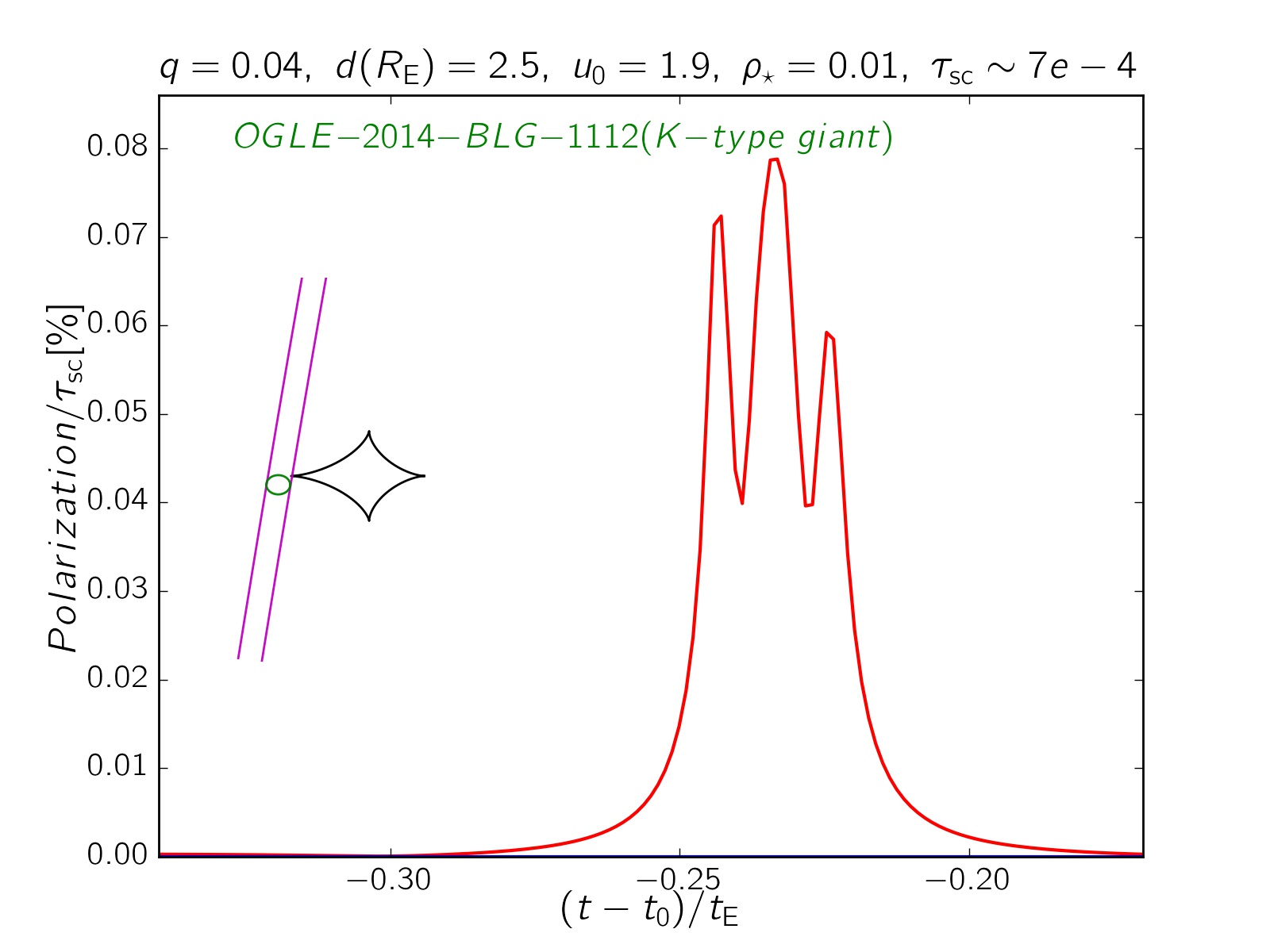}\label{fig4f}}
\caption{The expected polarization curves of six observed
binary-lens microlensing events with caustic crossings. The OGLE
names of these events and the kinds of their sources are pointed out
inside each panel with green color and their parameters are given at
the top of each plot.}\label{real}
\end{figure*}
\subsection{Case Study} In this section, we investigate the
expected polarization curves for some caustic-crossing microlensing
events recently recorded by OGLE, MOA and KMTNet surveys. The source
stars of most microlensing events toward the Galactic bulge are
either main-sequence or RCG stars. For RCG sources, the local Stokes
intensities are given by integrating on the polarization source
functions over the line of sight coordinate
\citep{simmons2002,Ignace2006}. For these stars, the polarization
signal linearly depends on the scattering optical depth which is
given by:
\begin{eqnarray}
\tau_{\rm{sc}}=\frac{n_{0}\sigma R_{\rm{h}}}{\beta-1},
\end{eqnarray}
where $\sigma$ is the scattering cross section, $n_{0}$ indicates
the strength of scattering number density of gas and dust in the
stellar envelope $n$, which can be described by:
\begin{eqnarray}
n=n_{0}(\frac{R_{\rm{h}}}{r})^{\beta},
\end{eqnarray}
where $R_{\rm{h}}$ is the halo (inner) radius of this density
function, i.e., $n=0$ for $R<R_{\rm{h}}$. However, $n_{0}$ itself
depends on the mass loss, $R_{\rm{h}}$ and the wind terminal
velocity in the stellar atmosphere \citep{Ignace2008}.

For main-sequence stars, the limb-darkening coefficients strongly
depends on the stellar surface temperature which specifies the
contributions of Rayleigh and Thomson scatterings
\citep{Claret2004}. More details about their polarization profiles
can be found in \citet{Ingrosso2012}. Here, we use their models to
evaluate the expected polarization curves of six caustic-crossing
and confirmed microlensing events. These curves are represented in
Figure (\ref{real}).

The First microlensing event is
$\rm{OGLE}$-$\rm{2015}$-$\rm{BLG}$-$\rm{0232}$ which was
characterized by \citet{20150232}. It is a binary-lens microlensing
event of a $\rm{K}$-type and main-sequence star. In this event, the
source passes close to the location of the primary. At the time when
the source disk is shown with red circle, the polarization signal
reaches to $0.03\%$.

The second microlensing event is
$\rm{OGLE}$-$\rm{2016}$-$\rm{BLG}$-$\rm{1067}$ whose source is a
$\rm{G5}$-type main-sequence star \citep{161067}. In this event,
there is no caustic crossings and the source just travels close to
the planetary caustics. The maximum polarization signal, i.e.,
$0.0001\%$, is too low to be detected with nowadays polarimeters,
e.g., the FOcal Reducer and low dispersion Spectrograph (FORS2)
polarimeter at Very Large Telescope (VLT) telescope with the
polarimetry accuracy $\sim0.1\%$.

The next selected microlensing event is
$\rm{OGLE}$-$\rm{2016}$-$\rm{BLG}$-$\rm{0596}$, a high-magnification
binary microlensing event. This event was characterized by
\citet{20160596}. The source star is a $\rm{K}$-type main-sequence
star with the apparent magnitude $\rm{I}=21.5~\rm{mag}$. The highest
polarization peak makes when the source edge is tangential to the
fold. Noting that the time scale of polarization peaks,
$t_{\star}\sim1.4~\rm{hr}$, is short, because of small source size.

The fourth microlensing event is
$\rm{OGLE}$-$\rm{2017}$-$\rm{BLG}$-$\rm{0329}$ which was studied by
\citet{20170329}. The source star is a $\rm{K}$-type giant star with
the apparent magnitude $\rm{I}=15.8~\rm{mag}$ at the distance
$D_{\rm{s}}=8.62~\rm{kpc}$. Its absolute magnitude is
$M_{\rm{I}}\sim1~\rm{mag}$, so its scattering optical depth is
$\tau_{\rm{sc}}\sim0.0001$. \footnote{We estimate the $\rm{I}$-band
extinction in the direction of each microlensing event and at the
source distances as explained in \citet{Radek2019} and the
scattering optical depth according to Figure (3) of
\citet{Ingrosso2015}}. The vertical axis of this plot is the
expected polarization signal normalized to the scattering optical
depth. In each fold crossing three polarization peaks produce which
are obvious in this plot.

The next event is a binary-lens microlensing of a red giant source,
$\rm{OGLE}$-$\rm{2017}$-$\rm{BLG}$-$\rm{0019}$. We use the best
fitted model recorded by V. Bozza in $\rm{RT}$-model
\footnote{\url{http://www.fisica.unisa.it/gravitationAstrophysics/RTModel/2017/OB170019_X1.pdf}}.
The source star should be a red giant, because of its large size,
i.e., $\rho_{\star}=0.02$. Its apparent magnitude is
$\rm{I}=14.8~\rm{mag}$ and the scattering optical depth is
$\sim4e-4$.

The last event is $\rm{OGLE}$-$\rm{2014}$-$\rm{BLG}$-$\rm{1112}$.
This event is a binary microlensing event of a $\rm{K}$-type giant
star \citep{141112}. The apparent magnitude of the source star is
$\rm{I}=13.98~\rm{mag}$. The absolute magnitude of the source star
is $M_{\rm{I}}=-0.7~\rm{mag}$ which results the scattering optical
depth $\tau_{\rm{sc}}\sim7e-4$. In this event the source passes very
close to the cusp but does not cross it.

Accordingly, most of caustic crossing microlensing events toward the
Galactic bulge whose source stars are RGB or main-sequence do not
produce enough high polarization signal to be detected with nowadays
polarimeters.

\section{Summary and conclusions}\label{four}
In this work, we reconsidered the polarization in binary-lens and
caustic-crossing microlensing events. In this regard, we introduced
the polarization maps which indicated the polarization signal of a
source star at every given position projected on the lens plane.
These maps display where the polarization signals maximize in
binary-lens microlensing events which in turn helps to find the best
candidates of on-going binary microlensing events for polarimetry
follow-up observation in near future.

According to different polarization maps, when the source radius is
sufficiently smaller than the caustic, the highest polarization
signal happens when the source is passing the folds which meet at
the nearest cusp to the location of the primary. However, while fold
caustic crossings three polarization peaks take place at the times
when the source edge is tangential to the fold (while fold caustic
crossings it happens two times) and when the source center is on the
fold. The highest one is when the source edge is upon the fold and
its center is out of the caustic. Noting that the maximum
magnification signals happen when the source center crosses the
on-axis cusp close to the primary.

On the connection line between caustic curves the polarization is
considerable, although it is not as high as while the fold caustic
crossings. When a source star is crossing this line, three
polarization peaks form. While the source passes normal to it,
(i)the largest peak takes place when the source center is upon the
connection line, (ii) the polarization angle at the time of the
largest peak differs by $90^{\circ}$ from those at the times of
other peaks.

When the source is quite into the caustic curve, there is no
considerable polarization signal, although the magnification factor
is high. In that case, all points of the source surface are
magnified and there is no remarkable symmetry breaking.

When the source size is on the order of the caustic or larger than
it, the locus of the source positions with highest polarization
signals ($\geq 0.6\%$) make a circular ring whose center is on the
primary's position. Its radius is somewhat smaller than the source
radius ($\sim 0.96~\rho_{\star}$). The larger source radius, the
higher polarization signal over this ring.

While cusp caustic crossings the behavior of polarization curves
depends on the source trajectory with respect to the symmetric axis
of cusps. Generally, while cusp crossings there are three
polarization peaks, the main (largest and widest) peak always
happens when the source center is into the caustic curve and its
edge is on the corner of the cusp. One (or two) smaller and narrower
peak(s) forms when the source is into the caustic and its edge(s) is
tangential to the folds meet at the cusp. When the source passes on
the symmetry axis of the cusp, there is a very small peak at the
time that the source is completely is out of the caustic and its
last edge is upon the corner of the cusp.

The time interval between the magnification peak and one of small
polarization peaks which appears at the first time that the source
is completely inside the caustic (two edges of the source are
tangential to the folds) is exactly $t_{\star}$.

\textbf{Acknowledgment} I thank R. Ignace and J. P. Harrington for
useful comments and also the anonymous Referee for his/her helpful
comments and suggestions. The work by the author was supported by a
grant (95843339) from the Iran National Science Foundation (INSF).

\bibliographystyle{mnras}
\bibliography{paperlast}

\appendix
\section{Photometry maps}
\begin{figure*}
\centering
\includegraphics[angle=0,width=0.495\textwidth,clip=0]{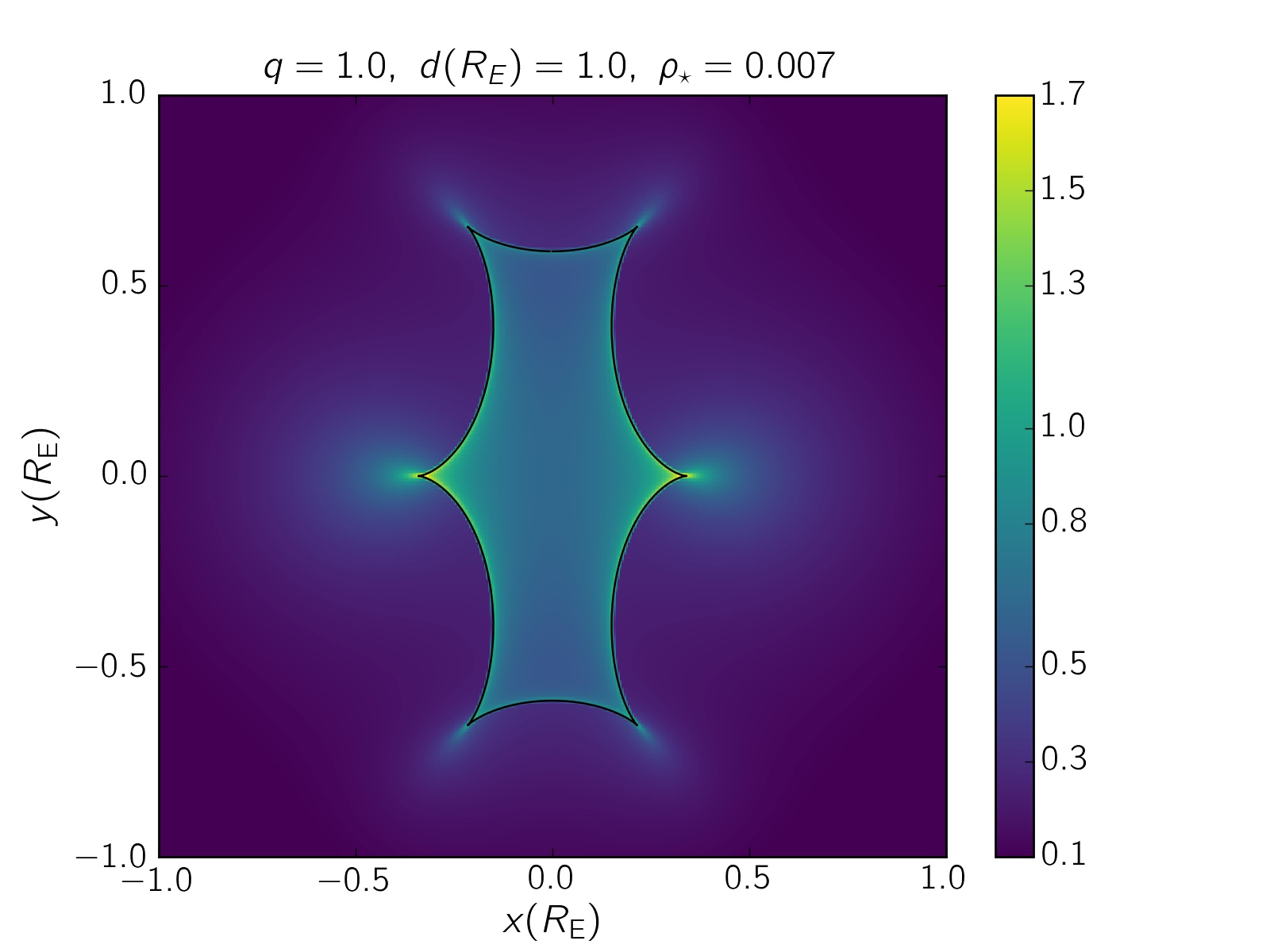}
\includegraphics[angle=0,width=0.495\textwidth,clip=0]{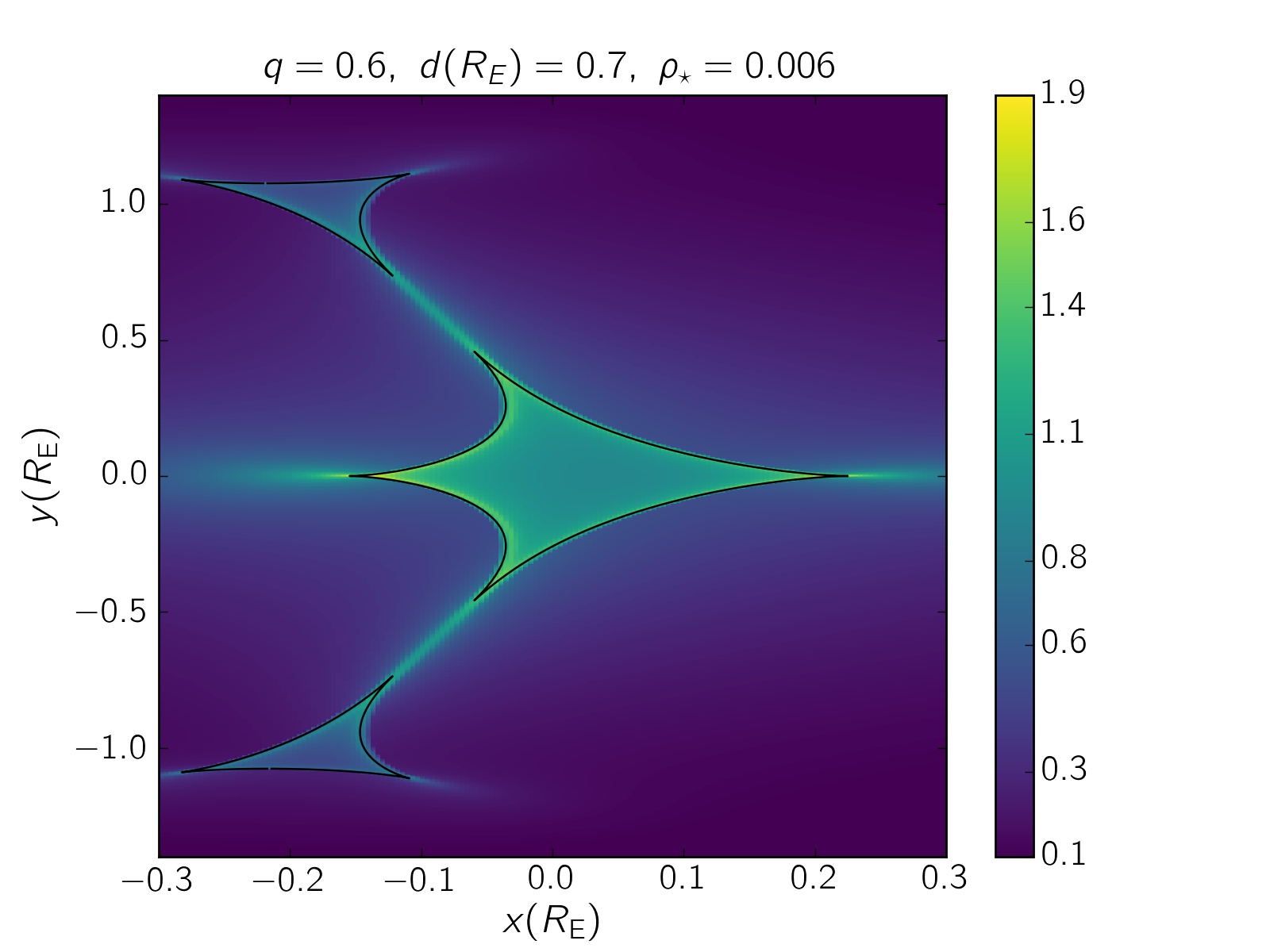}
\includegraphics[angle=0,width=0.495\textwidth,clip=0]{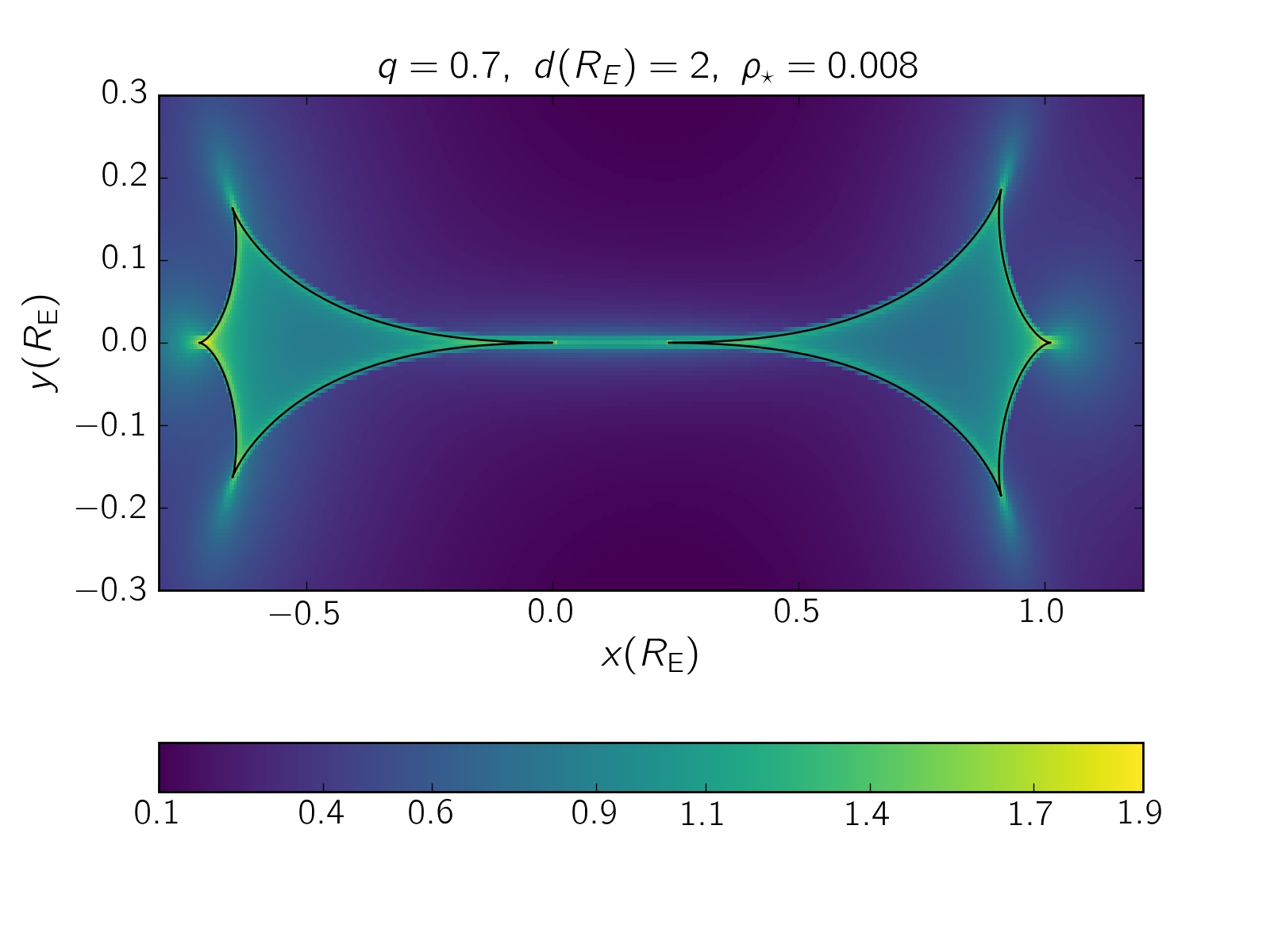}
\includegraphics[angle=0,width=0.495\textwidth,clip=0]{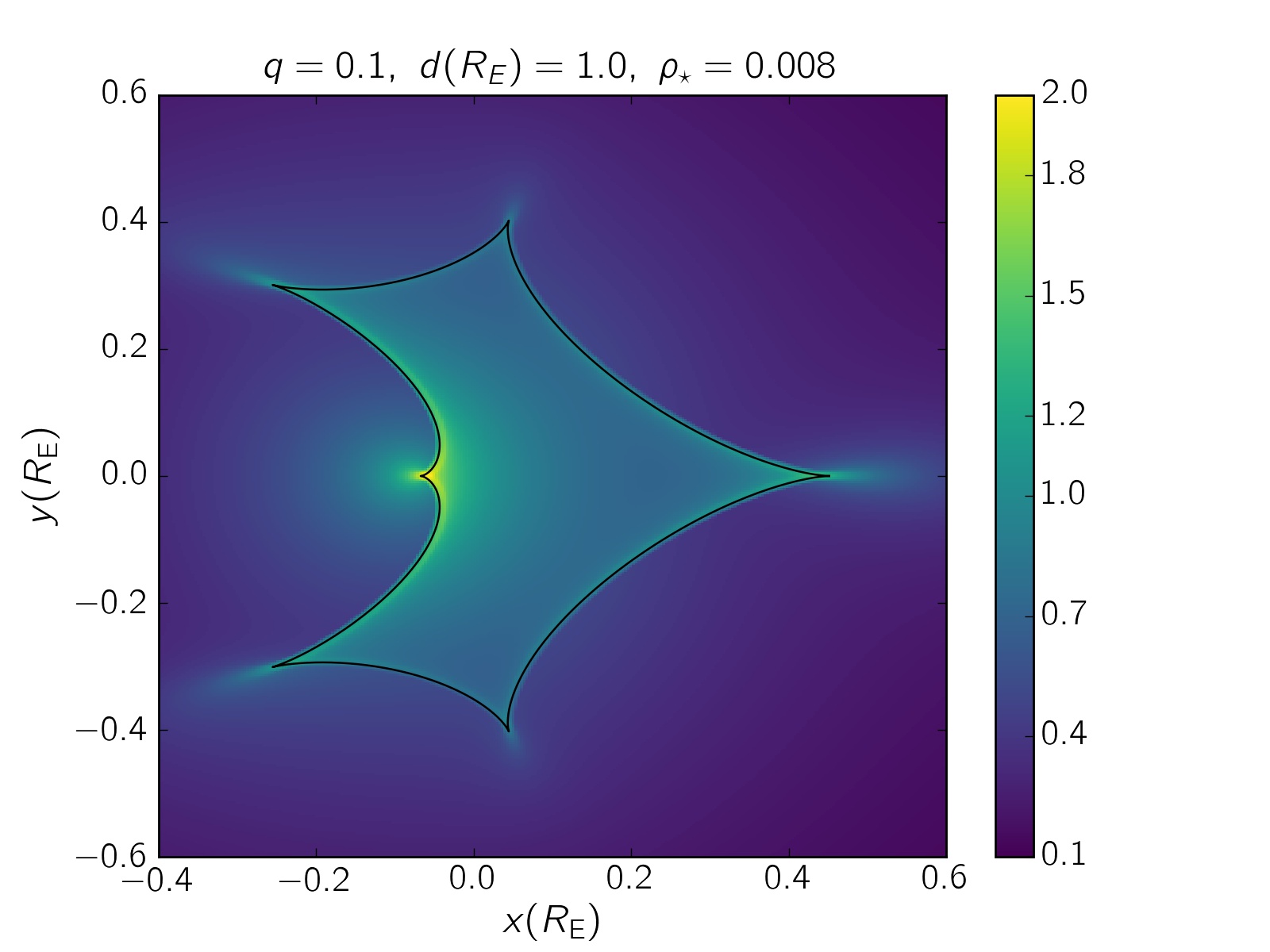}
\caption{Magnification maps over the lens plane for four different
binary-lens systems. The maps are in the logarithmic scale. The
corresponding polarization maps are plotted in Figure
(\ref{maps}).}\label{ap1}
\end{figure*}

\begin{figure*}
\centering
\includegraphics[angle=0,width=0.495\textwidth,clip=0]{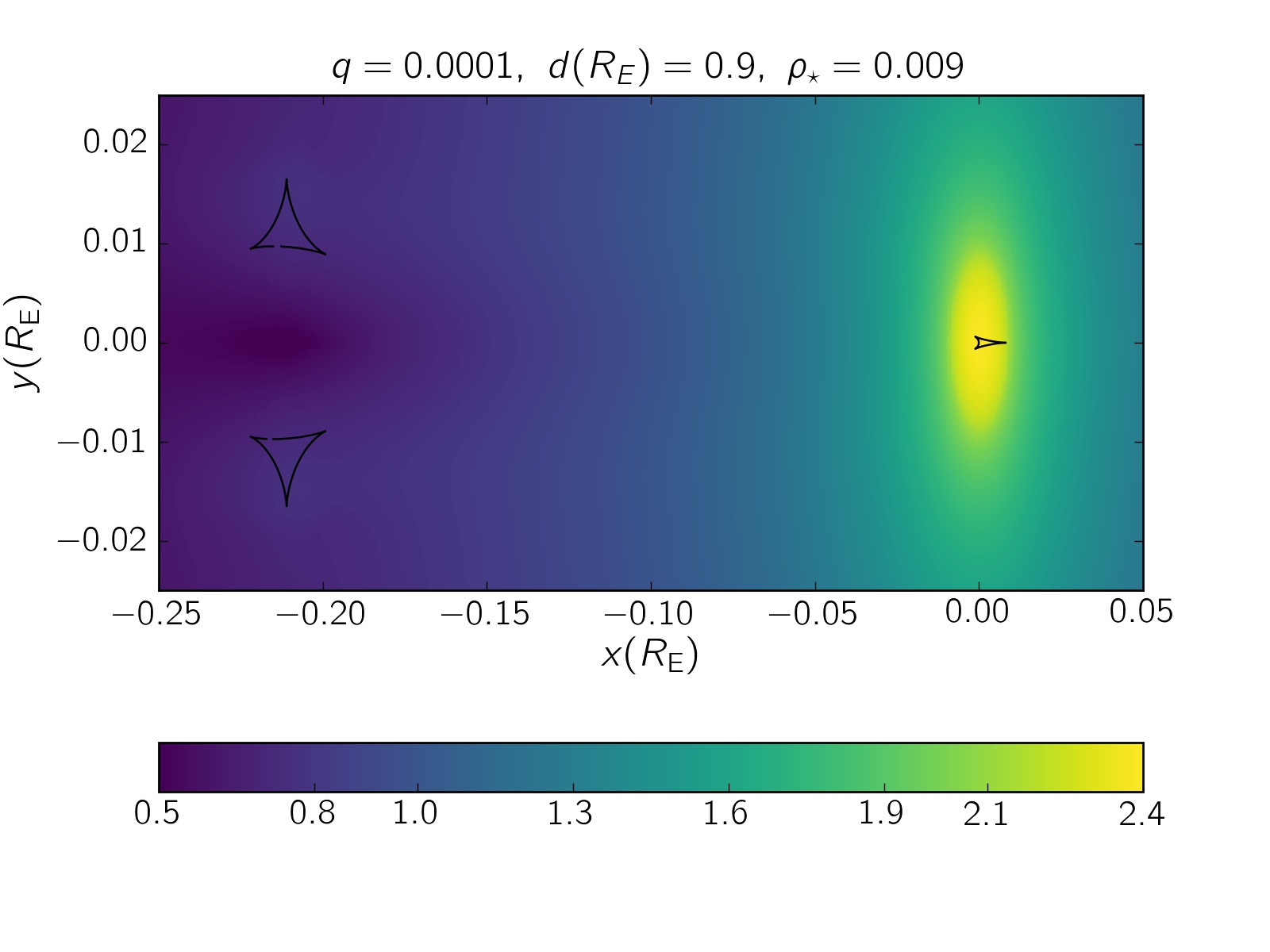}
\includegraphics[angle=0,width=0.495\textwidth,clip=0]{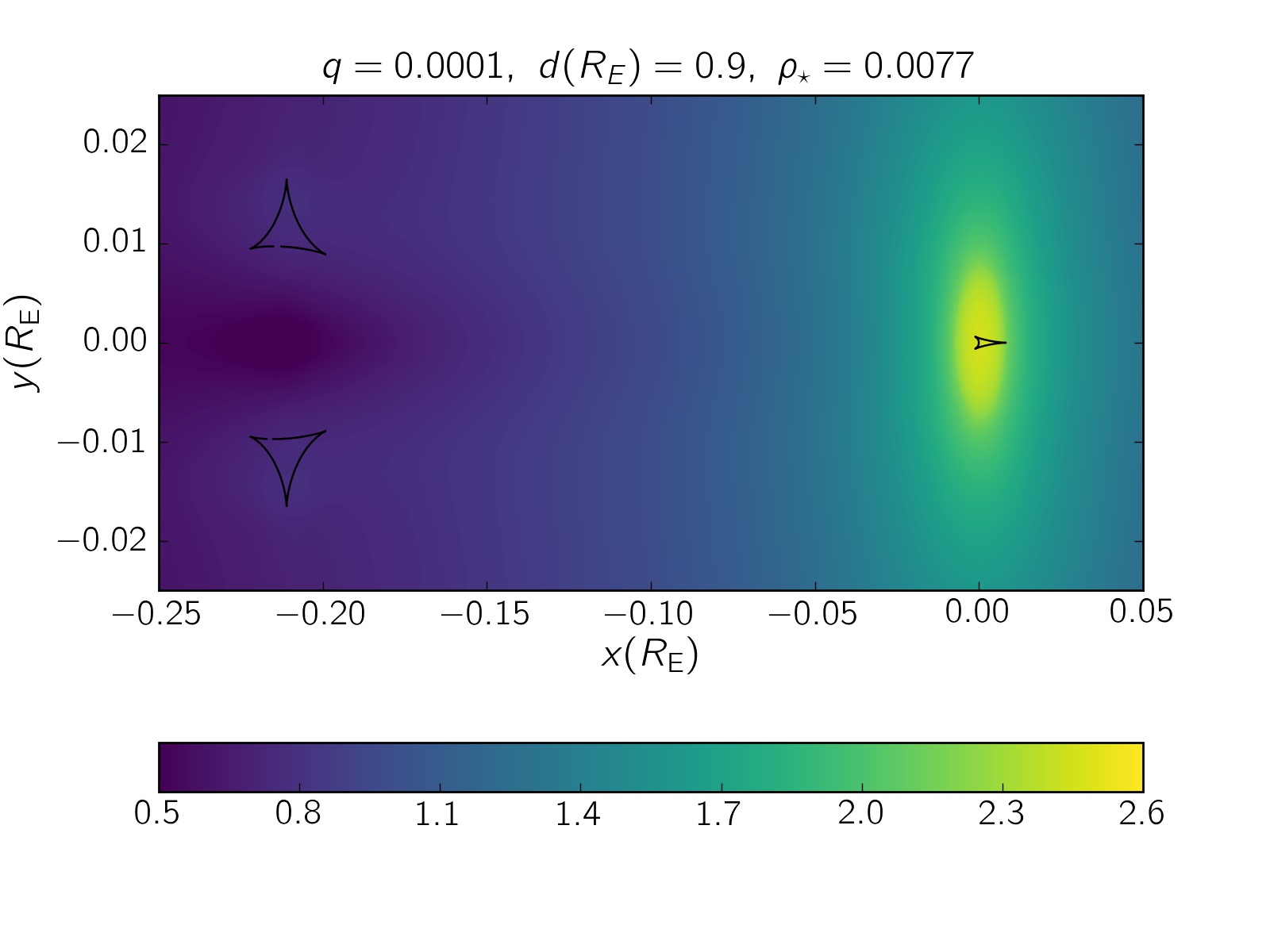}
\includegraphics[angle=0,width=0.495\textwidth,clip=0]{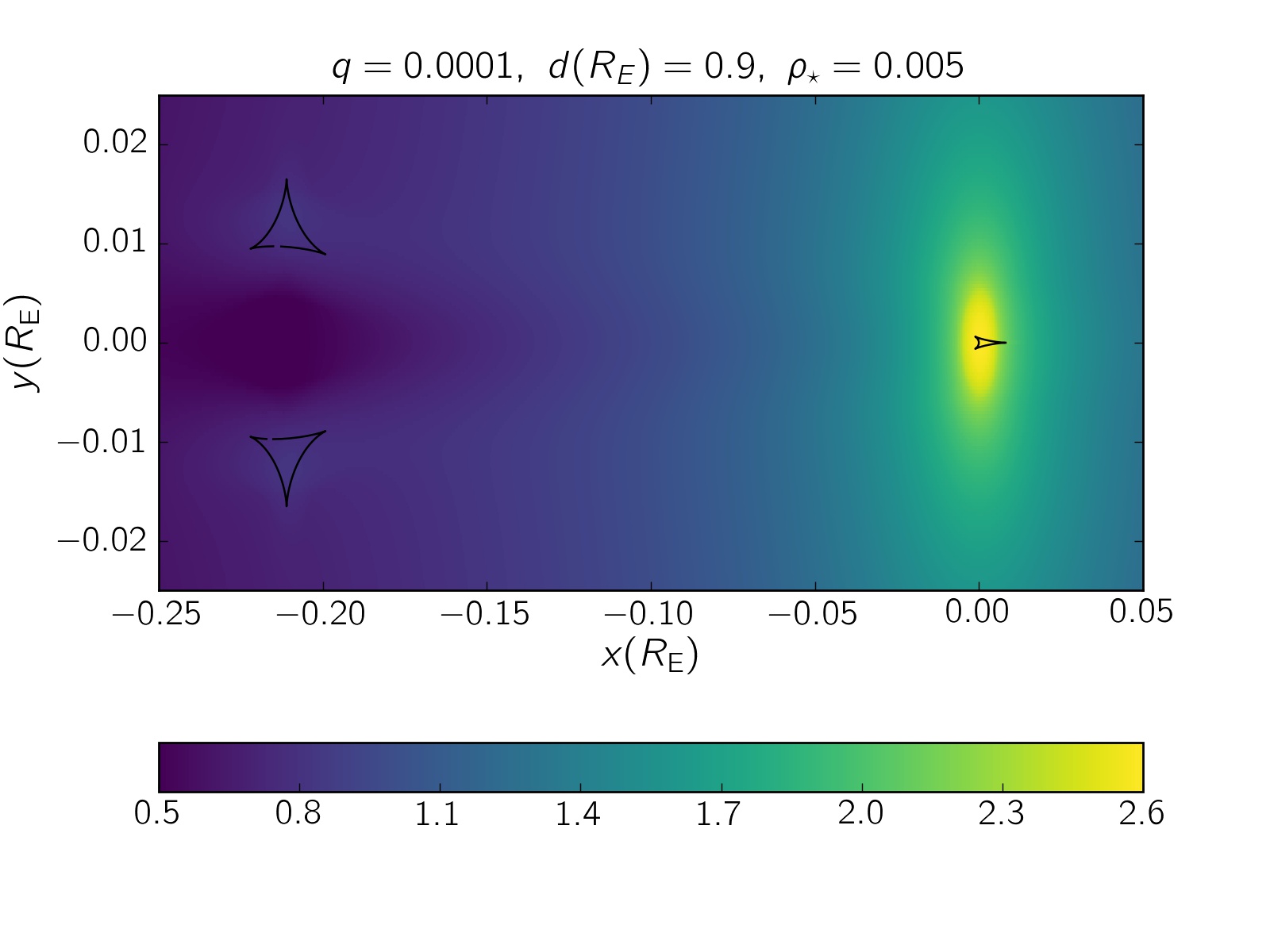}
\includegraphics[angle=0,width=0.495\textwidth,clip=0]{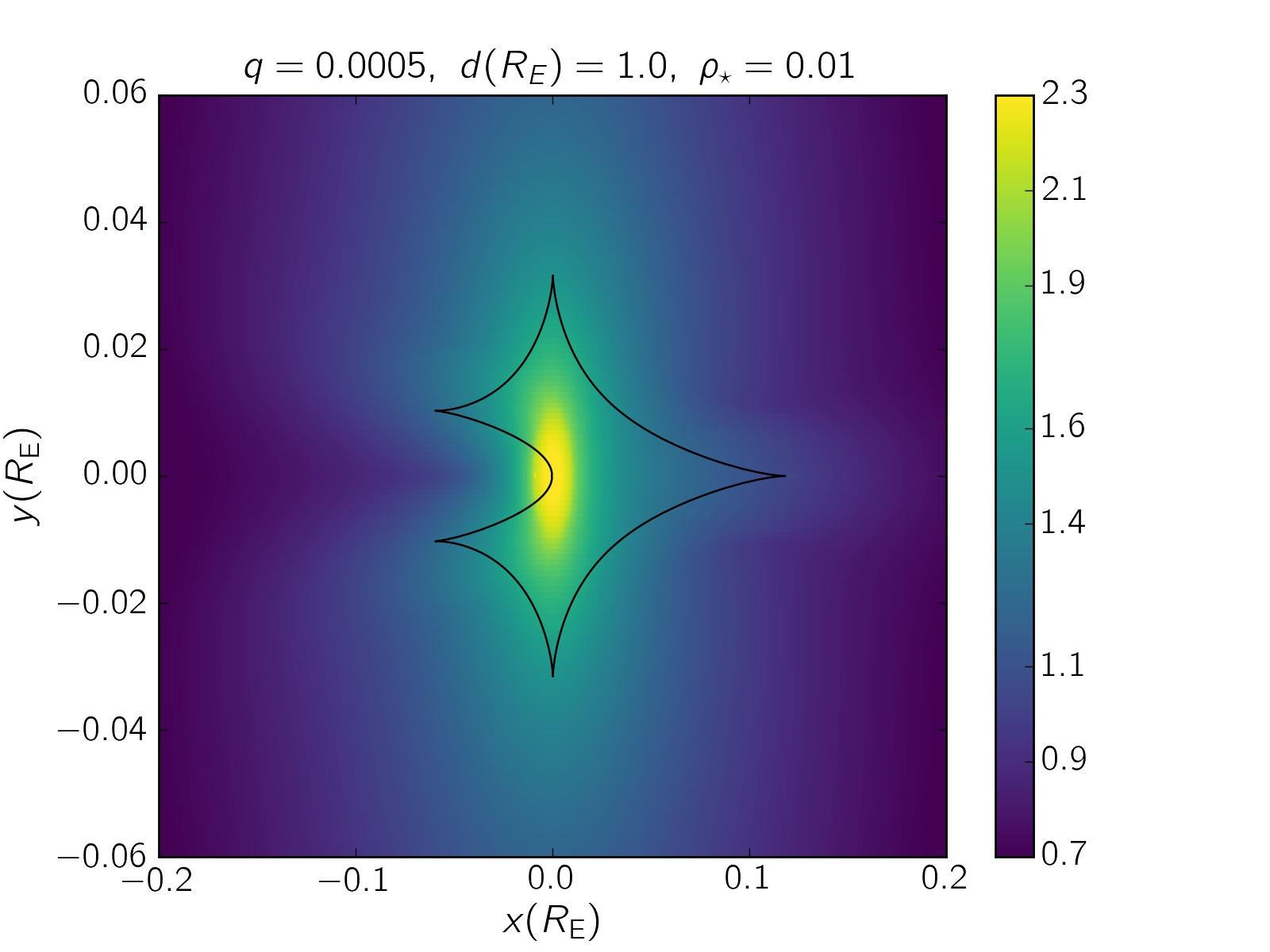}
\caption{Magnification maps over the lens plane for four
configurations of binary-lens systems and finite source effects. The
maps are in the logarithmic scale. The corresponding polarization
maps are plotted in Figure (\ref{planet}).}\label{ap2}
\end{figure*}

In order to compare the magnification and polarization behaviors in
binary-lens microlensing events, in Figures (\ref{ap1}) and
(\ref{ap2}) we represent the magnification maps on the lens planes
for different binary-lens configurations. Their corresponding
polarization maps are plotted in Figure (\ref{maps}) and
(\ref{planet}), respectively.
\bsp
\label{lastpage}
\end{document}